\documentclass[12pt, a4paper]{article}
\pdfoutput=1
\usepackage{graphicx}
\usepackage{amssymb}
\usepackage{amsmath}
\usepackage{bm}
\usepackage{color}
\usepackage{theorem}
\usepackage{subfigure}

\usepackage[sort&compress,numbers, merge]{natbib}

\definecolor{Orange}{cmyk}{0,0.61,0.87,0}
\definecolor{JungleGreen}{cmyk}{0.99,0,0.52,0}
\definecolor{OliveGreen}{cmyk}{0.64,0,0.95,0.40}
\definecolor{Brown}{cmyk}{0,0.81,1,0.60}
\definecolor{RoyalBlue}{cmyk}{0.71,0.53,0,0.12}

\usepackage[colorlinks=true, linkcolor=OliveGreen, citecolor=RoyalBlue,
urlcolor=Brown]{hyperref}

\begin{document}

\begin{titlepage}

\begin{flushright}
KIAS--P16015 \\
FTPI--MINN--16/04 \\
UMN--TH--3514/16
\end{flushright}

\vskip 1.35cm
\begin{center}

{\LARGE
{\bf
Naturalizing Supersymmetry with a\\[3pt]
Two-Field Relaxion Mechanism
}
}

\vskip 1.5cm

Jason L. Evans$^a$,
Tony Gherghetta$^b$,
Natsumi Nagata$^c$,
and
Zachary Thomas$^c$

\vskip 0.5cm

{\it $^a$School of Physics, KIAS, Seoul 130-722, Korea} \\[3pt]
{\it $^b$School of Physics \& Astronomy, University of Minnesota,
 Minneapolis, MN 55455, USA}\\[3pt]
{\it $^c$William I. Fine Theoretical Physics Institute, School of
 Physics \& Astronomy, \\ University of Minnesota, Minneapolis, MN 55455,
 USA}

\date{\today}

\vskip 1.5cm

\begin{abstract}
 We present a supersymmetric version of a two-field relaxion model
 that naturalizes tuned versions of supersymmetry. This arises from a
 relaxion mechanism that does not depend on QCD dynamics and where the
 relaxion potential barrier height is controlled by a second axion-like
 field. During the cosmological evolution, the relaxion rolls with a nonzero
 value that breaks supersymmetry and scans the soft  supersymmetric mass terms.
 Electroweak symmetry is broken after the soft masses become of order the
 supersymmetric Higgs mass term and causes the relaxion to stop rolling for
 superpartner masses up to $\sim 10^9$ GeV. This can explain the tuning in
 supersymmetric models, including split-SUSY models, while preserving the
 QCD axion solution to the strong CP problem. Besides predicting two very
 weakly-coupled axion-like particles, the supersymmetric spectrum may
 contain an extra Goldstino, which could be a viable dark matter candidate.

\end{abstract}

\end{center}
\end{titlepage}

\section{Introduction}

A radically new idea that attempts to address the electroweak hierarchy
problem via a dynamical relaxation mechanism~\cite{Abbott:1984qf} was
recently proposed in Ref.~\cite{Graham:2015cka} (see also
Refs.~\cite{Dvali:2003br, Dvali:2004tma, Kobakhidze:2015jya, Espinosa:2015eda, Hardy:2015laa,
Patil:2015oxa, Antipin:2015jia, Jaeckel:2015txa, Gupta:2015uea,
Batell:2015fma, Matsedonskyi:2015xta, Marzola:2015dia, Choi:2015fiu,
Kaplan:2015fuy, DiChiara:2015euo, Ibanez:2015fcv, Fonseca:2016eoo, Fowlie:2016jlx} for
related work). A slowly rolling axion-like field, the relaxion, is used
to neutralize the mass-squared parameter in the Higgs potential during a
prolonged de-Sitter phase of the Universe. After the Higgs mass-squared
parameter becomes negative and electroweak symmetry is broken, the QCD
quark condensate generates a new, stabilizing contribution to the
relaxion potential, which stops the relaxion at a value corresponding to
a hierarchically small Higgs potential minimum. By carefully choosing
the slope of the relaxion potential, the correct electroweak vacuum
expectation value (VEV) can be obtained, thereby solving the hierarchy
problem in a technically natural way. However since in the minimal model
the relaxion is identified with the QCD axion \cite{Weinberg:1977ma,
Wilczek:1977pj} of the Peccei--Quinn mechanism \cite{Peccei:1977hh,
Peccei:1977ur}, the QCD $\theta$ angle is
no longer small. This can be avoided by adding ad hoc inflaton-relaxion
couplings which prevents a large $\theta$ angle. Alternatively one can
decouple the strong CP problem from the relaxion mechanism by
introducing a separate technicolor sector with new fermions that mimic
the QCD quarks and generate the relaxion stopping potential. While this
does address the $\theta$ problem, the new strong dynamics must be near
the electroweak scale, leading to a coincidence problem.

A more appealing, though non-minimal  scenario which addresses these
issues was considered in Ref.~\cite{Espinosa:2015eda}. The main feature
relies on generating a relaxion stopping potential that does not require
an additional source of electroweak symmetry breaking (such as, from QCD
in the minimal model). This allows the cutoff scale to be
significantly increased beyond the TeV scale up to $10^9$ GeV and since
it does not rely on QCD dynamics, the axion solution to the strong CP
problem is preserved. This is accomplished by generating an electroweak
invariant coupling to the relaxion periodic potential. However this term
also generates a large potential barrier via quantum corrections that
needs to be neutralized in order to allow the relaxion field to
roll. Hence besides the relaxion $\phi$, a second field, an
``amplitudon'' $\sigma$, must be introduced, which is responsible for
neutralizing the amplitude of the periodic potential, causing the
relaxion field to move. The coupled dynamics between $\phi$ and
$\sigma$ allows the relaxion to roll down, albeit discontinuously,
eventually stopping at a value with the correct electroweak VEV after the
Higgs mass-squared parameter becomes negative. This is again achieved by
technically natural small parameters in the relaxion
potential. Furthermore a crucial feature of the two-field model is that
the amplitudon, $\sigma$, must couple to a periodic function of $\phi$,
without a direct coupling to the Higgs field.

While the relaxion mechanism provides an alternative solution to the
hierarchy problem, it nonetheless may also play a crucial role in other
solutions to the hierarchy problem, such as supersymmetry (SUSY). Indeed a supersymmetric version of the original relaxion model was constructed in Ref.~\cite{Batell:2015fma}. A nonzero value of the relaxion now breaks supersymmetry and
as the relaxion rolls down the potential, it scans the soft masses.  When the soft mass scale becomes of order the supersymmetric mass $\mu$, the Higgs potential becomes unstable. Non-perturbative QCD dynamics then triggers a back reaction on the relaxion potential, causing the relaxion to stop rolling. While the Higgs field obtains its correct electroweak VEV and the soft mass parameters are of order $\mu$, there is no correlation between these two mass scales as in usual supersymmetric models.  This means that apparently tuned versions of supersymmetry with heavy sfermions can be naturally realized, thereby solving the little hierarchy problem in supersymmetric models.  However, since the model in Ref.~\cite{Batell:2015fma} is based on the original relaxion model, it also inherits the shortcomings of the nonsupersymmetric model.

Instead in this paper we construct a supersymmetric realization of the
two-field relaxion model presented in Ref.~\cite{Espinosa:2015eda}. The
relaxion is now part of a chiral superfield, $S$ that has a shift
symmetry corresponding to a global symmetry under which the matter and
Higgs fields are charged. The second field, $\sigma$ is the imaginary 
scalar component of a chiral superfield, $T$ that has a shift symmetry
from a different global symmetry under which the matter and Higgs field
are invariant. This then automatically forbids a coupling between
$\sigma$ and the Higgs field at the renormalizable level, which was a
crucial requirement in the non-supersymmetric model. Moreover our
supersymmetric model preserves the QCD axion solution to the strong CP
problem and eliminates any coincidences between the electroweak scale
and the scale of the non-perturbative dynamics. This is not unexpected
because these features are automatically inherited from the
non-supersymmetric model~\cite{Espinosa:2015eda}. 

The two real scalar fields in the relaxion sector undergo a classical,
slow-roll evolution during an inflationary phase of the Universe. Since
the periodic potential for the relaxion has a large amplitude,
conditions must also be satisfied by the second field $\sigma$, in order
to neutralize the barrier height and allow the relaxion to
roll. Eventually the relaxion must stop at a minimum that
corresponds to the correct electroweak Higgs VEV.  Together these
constraints will restrict the parameters in our model, which include the
shift-symmetry breaking parameters, the strong dynamics and global
symmetry breaking scales, and the scale of supersymmetry breaking.
In addition the inflationary sector must have a Hubble scale, $H_I$ that is 
at least compatible with $H_I \lesssim v$, where $v$ is the Higgs VEV, 
but will also be further restricted by the requirement of classical rolling.
We find that for typical values of our allowed parameters, sfermion mass
scales up to $10^9$ GeV can be accommodated for a range 
$10^{-18}\, {\rm GeV} \lesssim m_S \lesssim 10^{-4}\, {\rm GeV}$, 
where $m_S$ is shift-symmetry breaking mass parameter. Since our model can
tolerate much larger values of $m_S$ than that in Ref.~\cite{Batell:2015fma}, the 
minimum number of inflationary $e$-folds,
$N_e$ needed for the relaxion to complete its cosmological evolution can be reduced.
For $m_S \simeq 10^{-7}$ GeV and $H_I \simeq 1$~GeV, we obtain $N_e\gtrsim 10^{14}$.
Furthermore, for some values of $m_S$ it is even possible for the relaxion to
only have sub-Planckian field excursions and therefore additional
mechanisms for generating super-Planckian field values are not needed.
The constraints in our model are ameliorated compared to the
supersymmetric model considered in Ref.~\cite{Batell:2015fma}, primarily
due to the fact that the mass scales in the strongly-coupled sector
responsible for generating the relaxion periodic potential are not tied
to the electroweak scale.

The allowed sfermion mass scale, ranging up to $10^9$ GeV, helps to explain the apparent tuning in supersymmetric models arising from the little hierarchy between the electroweak and superpartner mass scales. In the case when the mediation scale is of order the global symmetry breaking scale, the soft mass parameters can be chosen to be of order the PeV scale ($10^6$ GeV), while the gaugino mass parameters are one-loop (or possibly further) suppressed. While this is similar to split-SUSY models~\cite{Wells:2003tf, ArkaniHamed:2004fb, Giudice:2004tc, ArkaniHamed:2004yi, Wells:2004di}, the $A$-terms are not necessarily loop suppressed, and may allow for a lower sfermion mass scale. In addition for super-Planckian field values of the relaxion, the relaxino (the fermionic superpartner of the relaxion) is the lightest supersymmetric particle. This is due to the fact that supergravity effects require an additional source of supersymmetry breaking that renders the gravitino heavy, but keeps the relaxino light. In this case the relaxino is no longer the Goldstino but could be a plausible dark matter candidate. Alternatively, if the mediation scale is much higher than the global symmetry breaking scale then the soft mass parameters are negligible at tree level and are instead induced radiatively from the gaugino masses. The low energy spectrum is similar to that of gaugino mediation~\cite{Kaplan:1999ac, Chacko:1999mi}, and with minimal variations can accommodate soft masses at the TeV scale. Thus, we see that the two-field relaxion mechanism can remove the tuning and solve the little hierarchy problem in a variety of supersymmetric models above the TeV scale.

Finally, a number of important issues in the relaxion mechanism can also be addressed and incorporated into our supersymmetric model. For instance, there are regions of parameter space for which the relaxion idea relies on super-Planckian field values of the scalar fields. Recently ideas~\cite{Kaplan:2015fuy,Choi:2015fiu} using a large number of axion fields can be used to generate an effective decay constant which can be arbitrarily larger than the Planck scale. In this way our explicit shift-symmetry breaking parameters in the relaxion potential can be related to the underlying ultraviolet (UV) completion containing $N$ axion fields. This allows the relaxion to be realized as an axion, albeit one not related to the strong CP problem in QCD. In addition the UV framework allows all symmetries to be gauged, which prevents the expected violation of global symmetries from quantum gravity effects.

The outline of this paper is as follows. In Section~\ref{sec:model} we present a supersymmetric extension of the two-field relaxion mechanism. The shift symmetries are introduced in Section~\ref{sec:shiftsym}, with the corresponding superpotential and K\"ahler potential terms. Supersymmetry breaking effects and the sparticle spectrum are discussed in Section~\ref{sec:susybreaking}, including the mass spectrum of the relaxion sector.  The condition for electroweak symmetry breaking is discussed in Section~\ref{sec:ewsb} and the generation of the relaxion periodic potential is presented in Section~\ref{sec:periodicpot}. The effects of supergravity are considered in Section~\ref{sec:supergravity}, which address the cosmological constant and the role played by the relaxino, and the post-evolution relaxion-sector mass spectrum is discussed in Section~\ref{sec:relaxionmass}.
In Section~\ref{sec:susyrelaxion} we determine the conditions required
for the supersymmetric relaxion mechanism. The conditions on the
dynamical evolution of the two scalar fields is considered in
Section~\ref{sec:scalarconditions} and the constraints for addressing
the little hierarchy problem are derived in
Section~\ref{sec:littlehierarchy}. Our concluding remarks are given in
Section~\ref{sec:conclusion}. In Appendix~\ref{sec:formulae}, we
summarize the Lagrangian terms in our model and present the detailed expressions 
for the supersymmetry breaking parameters. In Appendix~\ref{sec:uvcomp}, we present a UV completion of our effective 
low-energy model that provides an origin for the parameters in the
relaxion potential by introducing multi-axion fields.

\section{A Two-field Supersymmetric Relaxion Model}
\label{sec:model}

We begin by constructing a supersymmetric extension of the two-field
relaxion model given in Ref.~\cite{Espinosa:2015eda}, which contains two
real scalar fields, the relaxion, $\phi$ and the amplitudon,
$\sigma$. The complete Lagrangian of our model is presented in
Appendix~\ref{sec:formulae}. These two fields are embedded into two
Standard Model (SM) singlet chiral superfields, $S$ and $T$: 
\begin{align}
 S &= \frac{s + i\phi}{\sqrt{2}} + \sqrt{2}\, \widetilde{\phi}\, \theta + F_S \theta \theta ~, \\
 T &= \frac{\tau + i\sigma}{\sqrt{2}} + \sqrt{2}\, \widetilde{\sigma} \,\theta + F_T \theta \theta ~,
\end{align}
where $s, \tau$ are real scalar fields, $\widetilde\phi, \widetilde\sigma$ are the fermionic partners and $F_{S,T}$ are the auxiliary fields. Specifically, the relaxion field $\phi$ is identified with the imaginary scalar field component of $S$ and the amplitudon, $\sigma$ is identified with the imaginary scalar field component of $T$. The $\phi, \sigma$, fields will play a similar role to those considered in Ref.~\cite{Espinosa:2015eda}.

\subsection{Shift symmetries and Supersymmetry}
\label{sec:shiftsym}

The $\phi$ and $\sigma$ fields are assumed to transform under some global shift symmetry, which appears at the supersymmetric level as a shift symmetry on the superfields, $S$ and $T$, whereas the real scalar fields, $s$ and $\tau$ will remain unchanged. The shift symmetry ${\cal S}_S$ that keeps $\phi$ massless has the following transformation properties:
\begin{align}
 {\cal S}_S:~S&\to S + i\alpha f_\phi ~, \nonumber \\
T&\to T ~, \nonumber \\
Q_i &\to e^{iq_i \alpha} Q_i ~, \nonumber \\
 H_u H_d &\to e^{iq_H\alpha} H_u H_d ~,
 \label{Strans}
\end{align}
with $q_H \equiv q_{H_u} + q_{H_d}$, while the masslessness of $\sigma$ is preserved by imposing another shift symmetry, ${\cal S}_T$, with the fields transforming as:
\begin{align}
 {\cal S}_T:~S&\to S ~, \nonumber \\
T&\to T + i\beta f_\sigma ~, \nonumber \\
Q_i &\to  Q_i~,  \nonumber \\
 H_u H_d &\to  H_u H_d ~,
 \label{Ttrans}
\end{align}
where $\alpha,\beta$ are arbitrary constants. The minimal supersymmetric
Standard Model (MSSM) matter superfields are denoted by $Q_i$, while
$H_u$ and $H_d$ denote the MSSM Higgs superfields. As seen in
Appendix~\ref{sec:uvcomp}, these symmetries may originate from U(1)
symmetries in a UV theory. In this case, the decay constants $f_\phi$ and $f_\sigma$ are associated with the corresponding global symmetry breaking scales for which $\phi$ and $\sigma$ are respectively, the Nambu-Goldstone bosons. Since we are only interested in the effects of the Nambu-Goldstone bosons, we will only consider the effective theory below the scales $f_\phi$ and $f_\sigma$.

If the shift symmetries ${\cal S}_S$ and ${\cal S}_T$ were exact, then
the potential for both $\phi$ and $\sigma$ would completely vanish.
Since the relaxion mechanism relies on the cosmological evolution of
both of these fields, a nonzero potential is needed and the shift symmetry must be broken.
Therefore, we assume that the shift symmetries are softly broken by some
small parameter in the effective theory.  The origin of this small
breaking can be attributed to monodromy~\cite{Silverstein:2008sg,
McAllister:2008hb, Kaloper:2008fb, Flauger:2009ab, Kaloper:2011jz,
McAllister:2014mpa, Franco:2014hsa, Blumenhagen:2014nba,
Hebecker:2014kva}, or an effectively large decay constant of a periodic
potential \cite{Choi:2015fiu, Kaplan:2015fuy, Fonseca:2016eoo}. Just like in Ref.~\cite{Batell:2015fma}, we will simply parameterize the effects of the small shift-symmetry breaking by the following
superpotential terms:
\begin{equation}
  W_{S, T} = \frac{1}{2} m_S S^2 +\frac{1}{2} m_T T^2 ~,
\label{eq:wst}
\end{equation}
where $m_{S,T}$ are mass parameters. A UV model that generates these effective terms is presented in Appendix~\ref{sec:uvcomp}.\footnote{As pointed out in Ref.~\cite{Gupta:2015uea}, if $S$ and $T$ are Nambu-Goldstone superfields, then the terms in Eq.~\eqref{eq:wst} are incompatible with discrete gauge symmetries associated with these fields. This difficulty can be avoided if one considers a periodic potential with an effectively large period as the source of these terms. A concrete realization of this possibility is given in Appendix~\ref{sec:uvcomp}. }

In order for the relaxion mechanism to work, a potential which depends on the Higgs field and stops the rolling relaxion field needs to be generated. This back-reaction can arise from the dynamics of a strongly-coupled gauge theory. In the original relaxion model~\cite{Graham:2015cka}, QCD dynamics played this role with the QCD axion identified as the relaxion.
However this leads to an excessively large $\theta_{\rm QCD}$, so we will instead assume that there is a new SU($N$) gauge interaction which becomes non-perturbative at a scale $\Lambda_{N}$.
This SU($N$) gauge sector contains a pair of SM singlet vector-like superfields, $N$ and $\bar{N}$, which are in the fundamental and anti-fundamental representations of SU($N$), respectively. When the SU($N$) interaction becomes strong, these fields will confine and the condensation of the fermions generates a periodic potential for the relaxion field.

The K\"{a}hler potential in our model is a function
of $S+S^*$, $T+T^*$, and
the fields $Q_i, H_u, H_d, N, \bar{N}$, so that it remains invariant under the transformations ${\cal S}_S$ and ${\cal
S}_T$. An immediate consequence of this symmetry is that although the superfield $T$ can couple to the MSSM Higgs fields in the K\"{a}hler potential, the imaginary part of its scalar component $\sigma$, has no direct renormalizable coupling with $H_u$ and $H_d$.
This crucial feature in the two-field relaxion model, which was simply assumed in \cite{Espinosa:2015eda}, is naturally realized in our supersymmetric model.
On the other hand, the relaxion field $\phi$ can directly couple to the MSSM Higgs
fields in the K\"{a}hler potential via terms like $U(S + S^*, T +
T^*) e^{-\frac{q_H S}{f_\phi}}H_u H_d$ where $U$ is an
arbitrary function of $S+S^*$ and $T+T^*$.

As for the MSSM superpotential, the Yukawa terms are taken to be invariant under the shift symmetries, while the $\mu$-term is modified to
\begin{equation}
  W_{\mu} = \mu_0 e^{-\frac{q_HS}{f_\phi}} H_u H_d ~,
\label{eq:wmu0}
\end{equation}
so that it also preserves the shift symmetries. In our effective theory the mass scale $\mu_0$ must satisfy $\mu_0 \lesssim f_\phi, f_\sigma$. We also have the gauge kinetic terms for both the SM and additional SU($N$) gauge interactions, each with an anomalous coupling of $S$ to the gauge fields
\begin{eqnarray}
{\cal L}\supset \int d^2\theta  \left( \frac{1}{2g_a^2}-i\frac{\Theta_a}{16\pi^2}-\frac{c_a S}{16\pi^2f_\phi}\right){\rm Tr}({\cal W}_a{\cal W}_a) +{\rm h.c.}\label{eq:GaugKin}
\end{eqnarray}
where $a$ sums over the appropriate groups of SM$\times {\rm SU}(N)$, $\Theta_a$ is the $\theta$-term and $c_a$ is a constant which depends on the UV completion of the model.

As discussed in Ref.~\cite{Espinosa:2015eda}, the two-field relaxion model requires that the amplitude of the periodic potential of $\phi$ has a particular dependence on $\phi$, $\sigma$, and the Higgs fields. This can be realized in our supersymmetric model by including the following superpotential terms:
\begin{align}
 W_N = m_N N \bar{N} +
ig_S S N \bar{N} +ig_T T N \bar{N} +
 \frac{\lambda}{M_L} H_u H_d N \bar{N}~,
\label{eq:wn}
\end{align}
where $m_N$ is a supersymmetric mass parameter, $g_{S,T}$ are dimensionless couplings and
$M_L$ is a UV scale at which this higher-dimensional operator is generated. For example, if we consider heavy vector-like lepton superfields which couple $N$ and $\bar{N}$ to $H_d$ and $H_u$, respectively, then $M_L$ will be of order the mass scale of the vector-like leptons.

Note also that the second and third terms in Eq.~\eqref{eq:wn}
explicitly break the shift symmetries ${\cal S}_S$, ${\cal S}_T$;
thus, provided the couplings $g_S$ and $g_T$ are small, the naturalness
of the model is not spoiled~\cite{'tHooft:1979bh}. This is similar to
the role of the parameters $m_S$ and $m_T$ in Eq.~\eqref{eq:wst}. An
estimate of the size of the couplings $g_S$, $g_T$ in Eq.~\eqref{eq:wn} can be
obtained by considering the higher-dimensional K\"ahler term, $(S+S^*) N
{\bar N}$ suppressed by a UV mass scale, and similarly for
$T$. The $F$-term of $S$ ($T$), $F_S\sim m_S \phi$ $(F_T\sim m_T \sigma)$,
generates a coupling of order $g_{S,T}\sim m_{S,T}/f_{\phi,\sigma}$ if
the higher-dimensional operator is induced at the scale of $f_{\phi,
\sigma}$. Note that such K\"{a}hler potential terms do not break the shift symmetry by themselves; in this case all resulting shift-symmetry breaking arises from $W_{S,T}$.  However, the generation of these operators are model dependent,
and these couplings may be much smaller, as occurs in the UV
model considered in Appendix~\ref{sec:uvcomp}.

\subsection{SUSY-breaking and the sparticle spectrum}
\label{sec:susybreaking}

\subsubsection{The MSSM sector}
\label{sec:mssm}

During the cosmological evolution the background fields $\phi$ and $\sigma$ will induce supersymmetry breaking, although the $\sigma$ contribution will disappear once $\sigma$ reaches its minimum ($\sigma = 0$). Interestingly, this will generate soft SUSY-breaking mass terms for the MSSM fields. The fact that supersymmetry is broken can be seen by noting that the auxiliary fields, $F_S$ and $F_T$, have non-zero values when the background fields $\phi$ and $\sigma$ are non-zero. These auxiliary fields are given by linear combinations of $m_S \phi$ and $m_T \sigma$, depending on how they mix in the K\"ahler potential. Since the
values of $\phi$ and $\sigma$ undergo large changes during their evolution, the $F$-terms will change significantly as well. This causes a scanning of the soft masses as the $F$-terms vary~\cite{Batell:2015fma}.

\begin{figure}[t]
\centering
\includegraphics[clip, width = 0.4 \textwidth]{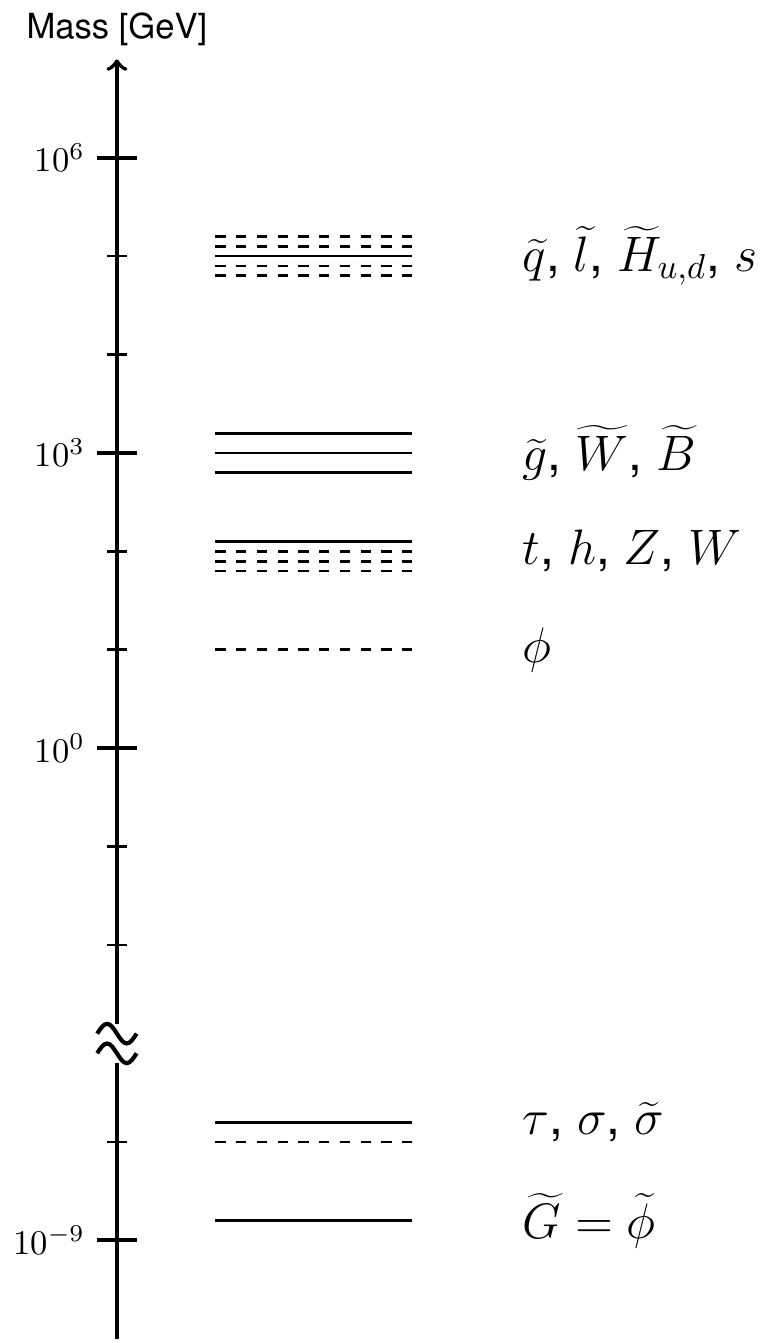}
\caption{A schematic diagram of an example mass spectrum in our model, where we have taken $m_{\text {SUSY}} =10^5$~GeV, 
$M_* \sim f_{\phi, \sigma}=10^5$~GeV, and $M_{ST} = M_P$. The MSSM mass spectrum is discussed in  Sec.~\ref{sec:susybreaking}, while the mass spectrum of the relaxion sector is discussed in Secs.~\ref{sec:supergravity} and \ref{sec:relaxionmass}.}
\label{fig:masssp}
\end{figure}

The soft SUSY-breaking parameters are readily obtained by expanding the K\"{a}hler potential and superpotential in terms of these $F$-terms (see Appendix \ref{sec:formulae}). This gives rise to two types of soft-mass spectra depending on the ``mediation'' scale in the K\"{a}hler potential. In the K\"{a}hler potential, $S$ and $T$ couple to matter fields through higher-dimensional operators suppressed by a messenger scale $M_*$. If $M_*$ is of order the decay constants $f_\phi$ and $f_\sigma$, then we expect that the soft masses, $\widetilde{m}_i$, the
scalar bilinear, $B$ and trilinear couplings, $A_{ijk}$, are ${\cal O}(F/M_*)$, where $F\sim F_{S,T}$. On the other hand gaugino masses, are suppressed by a loop factor compared with the soft masses since they are generated from the anomalous couplings of $S$ to the SM gauge fields and from gauge-mediation-like effects of $H_u$ and $H_d$. Therefore, this case realizes a split-SUSY-like mass spectrum \cite{Wells:2003tf, ArkaniHamed:2004fb, Giudice:2004tc, ArkaniHamed:2004yi, Wells:2004di} as discussed in Refs.~\cite{Batell:2015fma, Baryakhtar:2013wy}. Note, however, that the $A$-terms in this case are not suppressed by a loop factor contrary to other split-SUSY models based on anomaly mediation \cite{Randall:1998uk, Giudice:1998xp} (see, for instance,
Refs.~\cite{Ibe:2006de, Hall:2011jd, Ibe:2011aa, Ibe:2012hu,
Arvanitaki:2012ps, Hall:2012zp, ArkaniHamed:2012gw, Evans:2013lpa,
Evans:2013dza}). This difference can be phenomenologically important as
sizable $A$-terms may favor a lower SUSY-breaking scale for explaining
the observed Higgs mass \cite{Aad:2015zhl}. 
In Fig.~\ref{fig:masssp}, we show a typical example of the sparticle mass
spectrum for the $M_* \sim f_{\phi, \sigma}$ case, where we take 
$m_{\text{SUSY}}=f_{\phi, \sigma}=10^5$~GeV.

Moreover, the gaugino mass spectrum in our model can be different from that of
anomaly mediation; the gaugino masses in our model depend on the coefficients $c_a$ in
Eq.~\eqref{eq:GaugKin}, while those generated by anomaly mediation
are determined by the corresponding gauge coupling beta functions.
The bino and wino masses are also generated by the Higgsino-Higgs loop
diagram, similar to gauge mediation.
Note that $c_a$ can vanish if $q_i = q_H = 0$. In this case, gaugino masses, $M_a$
are induced by dimension-seven operators, such as $\int
d^4 \theta (S+S^*)^2 {\rm Tr}({\cal W}_a {\cal W}_a)/M_*^3$
\cite{ArkaniHamed:2004fb}, which yields, $M_a ={\cal O}
(|F|^2/M_*^3)$. This again gives rise to a split spectrum. In contrast to the anomaly mediation \cite{Randall:1998uk, Giudice:1998xp}
and the axion mediation \cite{Batell:2015fma, Baryakhtar:2013wy} scenarios,
where the gluino mass is large as a consequence of the size of the strong gauge
coupling beta function and the requirements of the Peccei--Quinn
mechanism, respectively, the gluino mass can be significantly suppressed. This situation may give an interesting signature at
the LHC; if the gluino is the lightest gaugino, it will decay into a
gravitino with a very long decay length. Such a gluino can be probed with displaced
vertex searches \cite{CMS:2014wda, Aad:2015rba} and $R$-hadron searches
\cite{Chatrchyan:2013oca, Aad:2015qfa}.

The second possibility is that the messenger scale $M_*$ is much higher than the 
decay constants $f_\phi$ and $f_\sigma$, such as the grand unification or Planck scale. 
In this case, $\widetilde{m}_i$ and $A_{ijk}$ are negligible at tree level but are induced radiatively 
from the gaugino masses, which are not suppressed by large $M_*$. Therefore, the 
low-energy SUSY spectrum becomes similar to that of gaugino mediation~\cite{Kaplan:1999ac, 
Chacko:1999mi, Schmaltz:2000gy, Schmaltz:2000ei, Gherghetta:2000kr, Csaki:2001em, Cheng:2001an} 
or no-scale models \cite{Cremmer:1983bf, Ellis:1983sf}. Furthermore, $S$
can be coupled to a vector-like pair of SM charged fields, {\it e.g.}, a
${\bf 5}$ and $\bar{\bf 5}$ of an SU($5$) grand unification, without
breaking ${\cal S}_S$, as was done for the Higgs bosons.\footnote{For
the UV completion in Appendix~\ref{sec:uvcomp}, we can couple $\phi_0$,
whose phase contains $S$, to this vector-like pair of ${\bf 5}$ and $\bar{\bf
5}$.}  When the ${\bf 5}$ and $\bar{\bf 5}$ are integrated out at the
scale $f_\phi$, they will induce the coupling between $S$ and the
gauge fields found in Eq.~\eqref{eq:GaugKin}, which generates the
gaugino masses.  As this is just a variation of gauge mediation (see
Ref.~\cite{Giudice:1998bp} and references therein), it will also generate
sfermion masses of similar size. In this scenario, the leptonic part of
the messengers can mix with the up-type Higgs superfield.\footnote{This
mixing can be realized in the UV completion in Appendix~\ref{sec:uvcomp}
through a superpotential term coupling between the multi-axion fields
$\phi_i$, with $H_u$ and the leptonic part of the $\bar{\bf 5}$.}  
This mixing generates $A$-terms which enhance the Higgs boson 
mass~\cite{Chacko:2001km,Evans:2011bea,Evans:2012hg,Kang:2012ra, Craig:2012xp,Abdullah:2012tq,Craig:2013wga}. 
This then allows the soft masses to be pushed down to the TeV scale. 
Although the tuning in this case is milder than in split-SUSY, it is still larger 
than expected and can be explained by the relaxion mechanism.

\subsubsection{The relaxion sector}
\label{sec:relaxion}

As we have mentioned, $\sigma$ does not couple to the MSSM Higgs fields in the SUSY-invariant Lagrangian terms. This feature is, however, violated by higher dimensional operators in the K\"ahler potential which communicate SUSY breaking to the Higgs fields. As discussed in Ref.~\cite{Espinosa:2015eda}, if the $\sigma$-Higgs couplings are sizable, then the late-time evolution of the $\sigma$ field will change the Higgs soft masses considerably, which makes it difficult to find a relaxion model that solves the little hierarchy problem. A simple way to avoid this is to assume $m_T \ll m_S$.
In fact, if $m_T$ were exactly zero then all of the $\sigma$-Higgs couplings would vanish\footnote{Even though $m_T = 0$, the (soft) mass of $\sigma$ can be induced at loop level via the coupling to $N$
and $\bar{N}$. } since they are generated via the $F$-terms of $S$ and $T$, which always has $\sigma$ multiplied by $m_T$.
Alternatively, the K\"{a}hler terms relevant to these couplings can be suppressed by some power of a large $M_*$. In this case, radiative effects of the gaugino masses, which depend only on $\phi$, become the dominant source for the soft mass parameters, and thus the $\sigma$ contribution to soft terms is negligible. Thus, in what follows we
will assume that these unwanted $\sigma$-Higgs couplings are sufficiently suppressed and the SUSY-breaking effects in the visible sector dominantly arise from the relaxion field $\phi$, rather than the amplitudon, $\sigma$.

Note that to determine the minimum of the potential with respect to $s$
and $\tau$, the fields $\phi$ and $\sigma$ can be regarded as
background fields. Because $\phi$ and $\sigma$ have very large field
values during their evolution,  we need only consider the potential at
leading order in $\phi$ and $\sigma$. The coefficients of these leading
order contributions are set by the components of the inverse K\"ahler
metric of $s$ and $\tau$, which is independent of $\phi$ and $\sigma$
due to the shift symmetry. If $S$ and $T$ mixing is present in the
K\"ahler potential, the solutions to the minimization conditions for
$s$ and $\tau$ will require a cancellation between terms with
coefficients of  $\sigma^2$, $\phi^2$ or $\sigma \phi$. Since $\sigma$
continues to evolve long after $\phi$ has settled into its minimum, the
location of the minimum will keep moving after the relaxion is
stabilized. This will lead to shifts in $s$ and $\tau$ of order
$f_\phi$. Since the size of $\mu_0$ is exponentially dependent on
$\frac{s}{f_\phi}$ for $q_H \neq 0$, changes in $s$ of this size will lead to order one
changes in the Higgs VEV. To prevent this from happening we must
require that there is no mixing between $S$ and $T$ in the K\"ahler
potential. This mixing can actually be significantly suppressed
as in the UV model considered in Appendix~\ref{sec:uvcomp}. In this
case, this mixing can be induced only by higher-dimensional operators. If
these operators are generated at a scale $M_{ST}$ much higher than $f_{\phi,
\sigma}$, then the off-diagonal components of the K\"ahler potential are
much smaller than the diagonal components, which results in negligible
mixing between $S$ and $T$.

The mass spectrum of the relaxion sector during the relaxation epoch is
determined by the K\"{a}hler and superpotential terms in this sector, and
thus is quite similar for the two SUSY-breaking scenarios discussed in
Sec.~\ref{sec:susybreaking}. The masses of $s$ and $\tau$ are ${\cal O}(|F_S|/f_\phi)$ and ${\cal
O}(|F_T|/f_\sigma)$ for both types of SUSY breaking, respectively,
as the mixing between $S$ and $T$ is negligible. Therefore,
the mass of $\tau$ is much smaller than the mass of $s$ when $|F_T| \ll
|F_S|$, which follows the requirement that the Higgs--$\sigma$ coupling
be negligible.

Finally, we note that the $N$ and $\bar{N}$ scalar components,
$\widetilde{N}$ and $\widetilde{\bar N}$, respectively, also appear in the scalar
potential of the relaxion sector via the superpotential couplings $g_S S
N \bar{N}$ and $g_T T N \bar{N}$. These fields are trapped at the origin during
the evolution of $\phi$ and $\sigma$ because of the $F$-term
contribution of $N$ and $\bar{N}$ to the scalar potential; this is
assured  as long as the couplings $g_S$ and $g_T$ times the field values
$\phi$ and $\sigma$ are larger than $m_S$ and $m_T$ in
Eq.~\eqref{eq:wst}, which can be easily realized thanks to the large
values of $\phi$ and $\sigma$, as we will see in
Sec.~\ref{sec:summaryplots}.

\subsection{Electroweak symmetry breaking}
\label{sec:ewsb}

Since we have determined the dependence of the SUSY parameters on $\phi$
and $\sigma$, we now examine the relation between the electroweak
symmetry breaking condition and the relaxion fields. This relation is a
crucial part of the relaxion mechanism. As discussed in
Section~\ref{sec:susybreaking}, the soft SUSY-breaking parameters are
assumed to be dominantly generated from $\phi$ and therefore the
electroweak symmetry breaking conditions will be expressed in terms of
$\phi$. The SUSY parameters relevant to the electroweak symmetry
breaking condition are parametrized as follows \cite{Batell:2015fma}:
\begin{align}
 m_{H_u}^2 &= c_u |m_S|^2 \phi^2,~~~~~~
 m_{H_d}^2 = c_d |m_S|^2 \phi^2,  \nonumber \\
 \mu &= c_{\mu 0} \mu_0 + c_\mu m_S^* \phi~, ~~~~~~
 B\mu = c_{B 0} \mu_0 m_S \phi + c_B |m_S|^2 \phi^2 ~,
\label{eq:EWsoftparam}
\end{align}
where the coefficients are model-dependent parameters.\footnote{{There
is also a periodic $B\mu$ term which causes a transition period during
which the determinant ${\cal D}(\phi)$ oscillates between positive and
negative values. For simplicity, we assume that this term is subdominant
with an amplitude $\lesssim v^2$.}} 
Note that $c_u$ and $c_d$
are real parameters, while the other parameters can be complex in
general. The order parameter for electroweak symmetry breaking is the
determinant of the Higgs mass matrix, which is a function of $\phi$: 
\begin{equation}
 {\cal D}(\phi) \equiv \left(m_{H_u}^2 +|\mu|^2\right)
\left(m_{H_d}^2+|\mu|^2\right) -|B \mu|^2 ~.
\label{eq:dphidef}
\end{equation}
We assume that initially $\phi$ has a very large field value and electroweak symmetry is preserved; namely,
\begin{equation}
 {\cal D}(\phi) > 0~~~~~\text{for}~~\phi \to \infty ~.
\end{equation}

As $\phi$ rolls down its potential, ${\cal D}(\phi)$ decreases. Eventually $\phi$ will reach a value for which the determinant becomes negative, triggering the breaking of electroweak symmetry. This critical value of $\phi$ is obtained by solving the equation, ${\cal D}(\phi) = 0$. Although there are analytic solutions of this quartic equation, we do not show them here because they are quite complicated. Instead, an order-of-magnitude estimate of the solution to ${\cal D}(\phi) = 0$ generically requires that
\begin{equation}
 \mu_0 \sim \frac{m_S \phi}{f_\phi} ~,
\label{eq:mu0msphi}
\end{equation}
where $\mu_0$ is the supersymmetric Higgs mass parameter in the superpotential and we have assumed that higher-dimensional terms which give rise to soft masses are suppressed by $M_*\sim f_\phi$ . Let $\phi_*$ denote the largest solution of ${\cal D}(\phi) = 0$ ({\it i.e.}, the value of $\phi$ at which ${\cal D}(\phi) = 0$ is satisfied for the first time during the cosmological evolution). Soon afterwards, $\phi$ stops rolling due to a Higgs-generated barrier in the periodic potential for
$\phi$ (see Sec.~\ref{sec:susyrelaxion}). At this point, the soft masses and the $\mu$ parameter\footnote{The Higgsino mass parameter $\mu$ can be different from the parameter in the superpotential, $\mu_0$, since there is also a SUSY-breaking contribution from the K\"{a}hler potential term $U(S + S^*, T + T^*) e^{-\frac{q_HS}{f_\phi}}H_u H_d$. } are given by
\begin{equation}
\mu \sim m_{\text{SUSY}}~, ~~~~
m_{H_u}^2 \sim m_{H_d}^2 \sim B\mu \sim m_{\text{SUSY}}^2 ~,
\label{eq:apprsoftmass}
\end{equation}
where $m_{\text{SUSY}}$ is a typical soft mass scale given by
\begin{equation}
 m_{\text{SUSY}}  \sim \mu_0 \sim \frac{m_S\phi_*}{f_\phi}~.
\label{eq:msusy}
\end{equation}
Therefore, the relaxion mechanism causes the soft masses to be of order the $\mu$ parameter scale,
yielding a solution to the little hierarchy problem.

In addition from Eq.~\eqref{eq:apprsoftmass}, we find that $\phi_*$ has a value:
\begin{equation}
 \phi_* \sim
10^{17}~{\rm GeV}\times
\left(\frac{m_{\rm SUSY}}{10^{5}~{\rm GeV}}\right)
\left(\frac{f_\phi}{10^{5} ~{\rm GeV}}\right)
\left(\frac{10^{-7}~{\rm GeV}}{m_S}\right)~.
\label{eq:phivalue}
\end{equation}
This shows that $\phi$, and thus also $\sigma$, can have sub-Planckian field values during the cosmological evolution.\,\footnote{In fact, this feature is also observed in the non-supersymmetric two-field relaxion model \cite{Espinosa:2015eda}. In this case, the lower bounds on the field values are given by $\phi \gtrsim \Lambda/g$ and $\sigma \gtrsim \Lambda/g_\sigma$, where these parameters are defined in Ref.~\cite{Espinosa:2015eda}. It is found that in a part of the allowed parameter region the fields $\phi$ and 
$\sigma$ can have sub-Planckian values without conflicting with these bounds.} However note that super-Planckian values are also possible which can lead to large effects from supergravity as we will discuss in Sec.~\ref{sec:supergravity}. 

When $M_* \gg f_\phi$, soft masses are induced by gaugino masses via
renormalization group effects. However, it turns out that the relations in
\eqref{eq:apprsoftmass} still hold in this case, while the right-hand
side of Eqs.~\eqref{eq:mu0msphi} and \eqref{eq:msusy} are suppressed by a
loop factor. This will lead to more stringent constraints on the parameter space.
In any case, the scale of the soft masses is determined by the 
supersymmetric Higgs mass parameter, $\mu_0$. Even if $\mu_0$ is much 
larger than the electroweak scale, the proper electroweak symmetry breaking 
conditions are realized
naturally thanks to the relaxion mechanism. Furthermore, note that in
this model the relaxion field $\phi$ scans the SUSY-breaking $F$-terms
and not the Higgs mass parameter directly. This feature is similar to
the model discussed in Ref.~\cite{Batell:2015fma}. For definiteness, we
focus on the $M_* \simeq f_\phi$ case in what follows.

As can be seen from Eq.~\eqref{eq:wmu0}, $\phi$ appears in the argument
of the $\mu$-term in the superpotential, and thus this framework
generically predicts an ${\cal O}(1)$ complex phase in the $\mu$
parameter. This phase can be probed by the measurements of electric
dipole moments, if the SUSY scale is ${\cal O}(1$--10)~TeV. For
instance, if winos are around the TeV scale and the Higgsino mass is not
so heavy, the two-loop Barr--Zee diagrams \cite{Barr:1990vd} induce a
sizable electron electric dipole moment \cite{Chang:2005ac, Deshpande:2005gi,
Giudice:2005rz}, which may be probed in future experiments.

In the case of interest when $m_{\text{SUSY}}$ is much higher than the electroweak scale, the effective theory for the Higgs sector below the SUSY-breaking scale contains a SM-like Higgs boson $h$ with a potential:
\begin{equation}
 V_{\text{Higgs}} (h) = \frac{1}{2} m_0^2 h^2
+\frac{1}{4}\lambda_h h^4~,
\end{equation}
where $m_0$ is a mass parameter and $\lambda_h$ is the quartic coupling. Since ${\cal D}(\phi)$ is the determinant of the Higgs mass matrix, the mass scale $m_0$ is related to the heavy Higgs mass $m_H$ by
\begin{equation}
 m_0^2 =  \frac{{\cal D}(\phi)}{m_H^2} ~,
\end{equation}
where $m_H ={\cal O} (m_{\text{SUSY}})$. The matching condition for the quartic coupling $\lambda_h$ is given at the SUSY breaking scale by
\begin{equation}
 \lambda_h = \frac{1}{8}(g^2 + g^{\prime 2}) \cos^2 2\beta  + \delta
  \lambda ~,
\end{equation}
where $g^\prime$ and $g$ are the U(1)$_Y$ and SU(2)$_L$ gauge coupling constants, respectively,  $\tan
\beta \equiv \langle H_u \rangle /\langle H_d \rangle$, and $\delta \lambda$ represents the threshold correction
that dominantly arises from top-stop loop diagrams. The value of the quartic coupling at the electroweak scale is obtained by renormalization group running (see, for instance, Refs.~\cite{Bagnaschi:2014rsa,
Vega:2015fna}). If ${\cal D}(\phi) < 0$, then $h$ acquires a VEV
\begin{equation}
 v(\phi) = \sqrt{-\frac{m_0^2}{\lambda_h}}
=\sqrt{-\frac{{\cal D}(\phi)}{\lambda_h m_H^2}} ~.
\label{eq:higgsvev}
\end{equation}
After the Higgs obtains a VEV, a back reaction on the relaxion potential will be responsible for stopping the rolling of the relaxion field.

\subsection{Generation of the periodic potential}
\label{sec:periodicpot}

In order for the relaxion mechanism to work, a back-reaction on the evolution of the relaxion field should occur immediately after electroweak symmetry breaking. In our model, the back-reaction arises from a relaxion-Higgs coupling in a periodic potential generated by a fermion condensate in a strongly-coupled SU($N$) gauge theory. To determine the amplitude of the periodic potential, we will first consider the Lagrangian for the fermionic components of $N$ and $\bar{N}$, denoted by $\psi_N$ and $\bar{\psi}_N$ respectively. These fermion terms arise from the superpotential $W_N$ in Eq.~\eqref{eq:wn}:
\begin{align}
 {\cal L}_N  =& -m_N \bar{\psi}_N \psi_N
- \frac{i}{\sqrt{2}} g_S (s + i\phi) \bar{\psi}_N \psi_N
-\frac{i}{\sqrt{2}}  g_T (\tau + i\sigma) \bar{\psi}_N \psi_N
-\frac{\lambda}{M_L} H_u H_d \bar{\psi}_N \psi_N
+\text{h.c.} \nonumber \\[3pt]
\simeq &\biggl[\left(
-\overline{m}_N +\frac{1}{\sqrt{2}}(\text{Re}(g_S) \phi+ \text{Re}(g_T) \sigma)
-\frac{\lambda}{M_L} H_u H_d
\right) \nonumber \\
&\quad +\frac{i}{\sqrt{2}}\left( \text{Im}(g_S) \phi
+\text{Im}(g_T) \sigma\right)
\biggr]\bar{\psi}_N \psi_N
+\text{h.c.} ~,
\end{align}
where we have used two-component Weyl notation and taken $\lambda$ to be
real. The mass parameter $\overline{m}_N \equiv m_N +\frac{i}{\sqrt{2}}
(g_S s + g_T \tau)$ is an effective mass of $\bar{\psi}_N \psi_N$
generated during the relaxation, and for simplicity we will assume this
parameter to be real.\footnote{The contribution of $g_S s + g_T \tau$ is
in fact negligible since we require that $\sigma$ monotonically decreases (see
Eq.~\eqref{eq:sigmadecmon}).} Under the transformation $\bar{\psi}_N
\psi_N\to e^{i\frac{c_a \phi}{\sqrt{2}f_\phi}}\bar{\psi}_N \psi_N$ the
term proportional to the $SU(N)$ $c_a$ in Eq. (\ref{eq:GaugKin}) 
disappears, generating a coupling between the fermion bilinear $\bar{\psi}_N \psi_N$
and the relaxion $\phi$.

After confinement the fermionic bilinear acquires a VEV $\langle\bar{\psi}_N \psi_N \rangle \sim \Lambda_N^3$, where $\Lambda_N$ is the confinement scale of the
SU($N$) gauge interaction. Assuming that $\Lambda_N \lesssim f_\phi,
f_\sigma$, this then gives rise to the following periodic
potential:\footnote{This ignores the gaugino condensate that would form
in a supersymmetric model.  This would make the amplitude $\mathcal{A}$
scale as $m_\lambda (m_N / \Lambda_N)^{1/N}$, which follows from an
analysis of the ADS superpotential \cite{Affleck:1983mk}.  The constraints and results of our analysis are essentially the same in this case, unless $N$ is very large. Alternatively, one can imagine that SUSY-breaking (either from $F_S$ or another SUSY-breaking sector such as that discussed in Sec. \ref{sec:supergravity}) raises the gaugino masses above $\Lambda_N$ and eliminates the gaugino condensate. Providing the necessary periodic potential with the gaugino condensate \emph{alone} is an intriguing possibility, but would require $\sigma$ to control the gaugino mass; this is difficult to accomplish without giving $\sigma$ an unwanted periodic potential.}
\begin{equation}
 V_{\text{period}} = {\cal A}\left(\phi, \sigma, H_uH_d\right) \Lambda_N^3
\cos \left(\frac{c_a \phi}{\sqrt{2}f_\phi}\right) ~,
\label{eq:periodicpot}
\end{equation}
where the amplitude ${\cal A}\left(\phi, \sigma, H_uH_d\right)$ is given by
\begin{equation}
 {\cal A}^2 \equiv \left[ \overline{m}_N -\frac{1}{\sqrt{2}}(\text{Re}(g_S) \phi + \text{Re}(g_T) \sigma)
+\frac{\lambda}{M_L} H_u H_d
\right]^2+\frac{1}{2}\left[\text{Im}(g_S) \phi
+\text{Im}(g_T) \sigma\right]^2 ~.
\end{equation}
The periodic potential (\ref{eq:periodicpot}) provides a barrier to the evolution of $\phi$ and therefore the amplitude ${\cal A}$ must vanish at some point during the cosmological evolution in order for $\phi$ to roll down its potential. Since electroweak symmetry is preserved during this epoch ($H_u=H_d=0$) the condition, ${\cal A} = 0$ implies
\begin{equation}
 \overline{m}_N = \frac{1}{\sqrt{2}}(\text{Re}(g_S)\phi + \text{Re}(g_T)\sigma)~~~\text{and}~~~
\text{Im}(g_S)\phi + \text{Im}(g_T) \sigma = 0~.
\end{equation}
These equations can be satisfied if both $g_S$ and $g_T$ are
real,\footnote{Alternatively ${\cal A} = 0$ can be realized if
$\overline{m}_N = 0$  and $g_S g_T^*$ is real. The latter condition
follows from $\text{Re}(g_S) \text{Im} (g_T)= \text{Re}(g_T) \text{Im}
(g_S)$ and is equivalent to $g_S,g_T$ having the same phase.} and
therefore we will assume this in what follows. After electroweak
symmetry is broken the $H_u H_d$ term in ${\cal A}$ will become sizable
and stop the relaxion field.

\subsection{Supergravity Effects}
\label{sec:supergravity}

As can be seen from Eq.~\eqref{eq:phivalue}, depending on the choice of
parameters, the relaxion, $\phi$ and $\sigma$ may undergo
super-Planckian field excursions.  In this case, one may rightly be
worried about supergravity effects dominating over the global SUSY
framework discussed above.  For example, take the scalar potential in
supergravity:
\begin{equation}
V = e^{K/M_P^2} \left(D^{i} W^*  D_i W  - \frac{3 |W|^2}{M_P^2} \right)~,
\end{equation}
where $M_P$ is the reduced Planck mass. Considering the potential for $\sigma$, the first term will give a positive contribution proportional to $m_T^2 \sigma^2$, while the second
gives a negative contribution proportional to $m_T^2 \sigma^4 / M_P^2$.
Since $\sigma$ is super-Planckian, the second term generically
dominates, regardless of the value of $m_T$ (or the vacuum expectation
value of $W$).  This gives a large negative contribution to the
cosmological constant at late times.\footnote{Presumably $\sigma$ does have a minimum rather than
a runaway direction; if one assumes the periodicity of $\sigma$ as in
Appendix~\ref{sec:uvcomp}, this is automatically realized.}
Therefore, some additional source of SUSY-breaking is needed in order to have a nearly-vanishing cosmological constant at the present time; note that contributions coming from $F_S$ may not be large enough if $\phi$ has super-Planckian excursions.  This potential is also very worrying due to its steepness; for a similar traversal of field values, the change in vacuum energy is enhanced by a factor of $(\Delta \phi / M_P)^2$, resulting in much more stringent constraints relating to the required Hubble scale of inflation.

The simplest way to arrange a vanishing cosmological constant at late times is to have (almost) no-scale SUSY breaking that is sequestered from the other sectors.  The resulting linear terms in the K\"{a}hler potential result in a cancellation of the $|W|^2$ term in the scalar potential, eliminating the problems mentioned in the previous paragraph.  Explicitly, this can be seen by expanding the scalar potential as:
\begin{equation}
V = e^{K/M_P^2} \left( W^{* i} W_i + \frac{1}{M_P^2} (W^{* i} K_i W + \textrm{h.c.})  + ( K^i K_i - 3 M_P^2 ) \frac{|W|^2}{M_P^4} \right)\,.
\label{eq:expandedscalarpotential}
\end{equation}
Almost no-scale models feature $K^i K_i \approx 3 M_P^2$ and $W_X \approx 0$ for the no-scale field $X$, so the latter two terms in Eq.~\eqref{eq:expandedscalarpotential} are suppressed relative to the first, resulting in a scalar potential almost identical to that of global SUSY.\footnote{The exponential prefactor does not have a relaxion contribution as long as the shift symmetries are not broken in the K\"{a}hler potential.}  However the latter two terms cannot generically be both eliminated altogether, since due to the SUSY-breaking in the relaxion sector, it is impossible to simultaneously satisfy
$K^i K_i = 3 M_P^2$ and $W_X = 0$, while maintaining a vanishing vacuum energy.  The resulting effects are suppressed by roughly $W_{S,T} / (F M_P)$, with $W_{S,T}$ the relaxion superpotential, and $F$ the overall SUSY-breaking order parameter (including the almost no-scale effects).  This is roughly equivalent to the condition that
\begin{equation}
\frac{F_S}{F} \lesssim \frac{M_P}{\Delta \phi}\,.
\label{eq:SUGRAconstraint}
\end{equation}
This condition is trivially satisfied, even without additional SUSY-breaking sectors, if the field excursion of $\phi$ is sub-Planckian.  Note also that even if this condition is violated, and the middle term in Eq.~\eqref{eq:expandedscalarpotential} were to dominate, it would have minimal effect on the relaxion mechanism itself, only rescaling a number of the constraints, as the overall relaxion potential would still be quadratic.\footnote{With a judicious choice of phase difference between the no-scale and relaxion superpotentials, the middle term can be made entirely imaginary in the relaxion sector and vanishes, providing another possible way to avoid this constraint. }

An almost no-scale structure, assuming it also existed while our relaxion fields were rolling (with additional SUSY-breaking contributions to provide inflation), also has the advantage of making the $F$-term of the conformal compensator (almost) vanish.  This ensures that the rolling of $\sigma$ and the resulting change in $W$ does not give sizable varying anomaly-mediated contributions to the Higgs soft terms after $\phi$ has stopped rolling.

These supergravity considerations do give one irreducible
phenomenological prediction. When most of the SUSY-breaking in the
universe does not come from $F_S$, the relaxino $\widetilde{\phi}$ is
not the Goldstino that is eaten by the gravitino in the super-Higgs
mechanism.  It is instead a second Goldstino, in the manner of
Ref.~\cite{Cheung:2010mc}.
In general, the gravitino mass $m_{3/2}$ is given by
\begin{equation}
 m_{3/2} = \frac{F}{\sqrt{3}M_P}
\simeq 2\times \left(\frac{F}{F_S}\right)
\left(\frac{m_{\rm SUSY}}{10^5~{\rm GeV}}\right)
\left(\frac{f_\phi}{10^5~{\rm GeV}}\right)~{\rm eV} ~.
\end{equation}
If the field excursion of $\phi$ is sub-Planckian, then the additional
SUSY-breaking contribution is not required, {\it i.e.}, $F=F_S$ is
allowed. In this case, $\widetilde{\phi}$ is the Goldstino eaten by the
gravitino, as discussed in Ref.~\cite{Batell:2015fma}.
As the additional SUSY-breaking effects increase, the gravitino
becomes heavier, and a larger part of the contribution to the helicity $\pm \frac{1}{2}$
part of the gravitino is provided by a Goldstino in the no-scale
sector. Note that in this case, the constraint of Eq. (\ref{eq:SUGRAconstraint})
implies that $m_{3/2} \gtrsim m_S (\Delta \phi / M_P)^2$.

The relaxino mass at tree level, on the other hand, takes on the value
\begin{equation}
m_{\widetilde{\phi}}  = - 2 m_{3/2} \frac{W_X}{F_X} \lesssim \frac{F_S}{M_P} \frac{F_S}{F_X} \lesssim m_S ~, \label{eq:relaxinomass}
\end{equation}
with $X$ the no-scale field, as can be calculated using the methods of
Ref.~\cite{Cheung:2011jq}. If it is not
too light, the relaxino $\widetilde{\phi}$ may be in the correct mass
range (keV to MeV) to be a plausible dark matter candidate.  As we take the relaxion sector to be the only source of SUSY-breaking, the relaxino effectively behaves like a gravitino in the early universe, except that it is lighter for a given coupling strength $F_S$ by at least a factor of $F_S/F_X$.  This can help to ameliorate the ``relaxino problem'' in the early universe.\footnote{Alternatively, an ${\cal
O}(1)$~eV gravitino does not suffer from the gravitino problem. }

Note that Eq.(\ref{eq:relaxinomass}) is strictly the mass of the uneaten Goldstino, $\it{i.e.}$ the field proportional to $F_X \widetilde{\phi} - F_S \widetilde{X}$.  In the sub-Planckian regime with $F_S \ll F_X$, this is the mass of the now-unnecessary no-scale fermion $\widetilde{X}$; the relaxino is eaten by the gravitino and has mass $m_{3/2}$.  Note that in the sub-Planckian regime, the last inequality in (\ref{eq:relaxinomass}) still holds, as $m_{3/2} \approx m_S \phi_* / M_P \lesssim m_S$.  The relaxino can still be dark matter in this regime, but it behaves as a conventional gravitino.

As an alternative, one can also
envision scenarios in which the no-scale sector is not strictly
sequestered, and gives substantial contributions to MSSM and
relaxion-sector soft masses.  These can substantially increase the
relaxino mass due to loop corrections, as in Ref.~\cite{Argurio:2011hs};
these are enhanced by the small $F_S/F$ required to satisfy
Eq.~\eqref{eq:SUGRAconstraint}.  Such further
effects are beyond the scope of this work.

\subsection{Post-evolution relaxion-sector mass spectrum}
\label{sec:relaxionmass}

At the end of the cosmological evolution, the relaxion is trapped at a local minimum induced by the periodic
potential, while the $\sigma$ field reaches its minimum at $\sigma=0$. Since these fields have settled in their
minima, the corresponding $F$-terms no longer change. The mass of $s$ is still ${\cal O}(|F_S|/f_\phi) \simeq
{\cal O}(m_{\rm SUSY})$, while the mass of $\tau$ is given by higher-dimensional operators that cause $S$--$T$
mixing in the K\"{a}hler potential, such as $\int d^4\theta (S+S^*)^2 (T+T^*)^2/M_{ST}^2$, and thus should be significantly
suppressed: ${\cal O}(m_{\rm SUSY} f_\phi /M_{ST})$. Regardless of the value of $M_{ST}$, the mass of $\tau$ is never below the supersymmetric mass $m_T$, barring a fine-tuning.

The relaxion mass, $m_\phi$, is rather large because of
the large amplitude of the periodic potential when $\sigma = 0$. It is
estimated as
\begin{align}
 m_\phi^2 &\simeq \frac{\Lambda_N^3}{f_\phi^2} {\cal A}
\left(\phi_*, 0, \frac{v^2(\phi_*)}{4} \sin 2\beta\right)
\simeq \frac{\Lambda_N^3 g_S\phi_*}{f_\phi^2}
\nonumber \\
&\simeq \left(\frac{g_S f_\phi}{m_S}\right)
\left(\frac{\Lambda_N}{f_\phi}\right)^3
\left(\frac{m_{\rm SUSY}}{f_\phi}\right) f_\phi^2 ~.
\end{align}
For example, if $g_S f_\phi/m_S = 10^{-8}$, and $m_{\rm SUSY} =
\Lambda_N = f_\phi = 10^5$~GeV, then the relaxion mass, $m_\phi \simeq 10$~GeV.
On the other hand, the $\sigma$ mass is simply the value of the supersymmetric mass, $m_T$; any corrections 
from SUSY breaking are subleading.

The mass of the relaxino $\widetilde{\phi}$ was discussed in the
previous section, but is always of order $m_S$ or less.  The mass of the
amplitudino $\widetilde{\sigma}$ will generically receive contributions
from K\"{a}hler potential terms such as $\int d^4 \theta
(S+S^*)(T+T^*)^2 / M_{ST}$ of order $m_S \phi_*/M_{ST}$.  These may or
may not dominate over the supersymmetric mass $m_T$.  For $\phi_*
\lesssim M_P$ and $M_{ST} \simeq M_P$, the Giudice-Masiero
contributions are of the same order as the relaxino mass and an
amplitudino LSP is possible if $m_T/m_S \lesssim \phi_*/M_P$; otherwise,
the relaxino is guaranteed to be the LSP.

A typical mass spectrum of the relaxion sector is shown in Fig.~\ref{fig:masssp}, where we
have assumed that $M_{ST} \simeq M_P$ and there is no additional SUSY breaking source. 
In this case, the relaxino $\widetilde{\phi}$ is eaten by the gravitino, and the masses of $\tau$,
$\sigma$, and $\widetilde{\sigma}$ are as large as the gravitino mass.~\footnote{Although our model has many light degrees of freedom, an extended period of inflation after the relaxion has settled into its minimum will suppress all vacuum energy in these fields.  Furthermore, the shift symmetries of the relaxion and amplitudino will lead to suppressed decays to these fields and reheating will in general produce very few of them.}

\section{The Supersymmetric Relaxion Mechanism}
\label{sec:susyrelaxion}

In order for the supersymmetric relaxion mechanism to solve the little hierarchy problem and obtain the correct Higgs VEV,
the model parameters must satisfy certain conditions during the cosmological evolution. The relevant potential terms for the evolution of $\phi$ and $\sigma$ are:
\begin{equation}
 V_{\phi,\sigma}(\phi, \sigma, H_u H_d)
=\frac{1}{2}|m_S|^2 \phi^2 +\frac{1}{2}|m_T|^2 \sigma^2
+{\cal A}\left(\phi, \sigma, H_u H_d\right) \Lambda^3_N
\cos\left(\frac{\phi}{f_\phi}\right) ~,
\label{eq:relaxionpotential}
\end{equation}
where we have taken $c_a=\sqrt{2}$ and
\begin{equation}
 {\cal A}\left(\phi, \sigma, H_u H_d\right)
=\overline{m}_N -\frac{g_S}{\sqrt{2}}\phi -\frac{g_T}{\sqrt{2}} \sigma
+\frac{\lambda}{M_L} H_u H_d ~.
\label{eq:simpamp}
\end{equation}
For simplicity, the K\"{a}hler
potential for $S$ and $T$ is assumed to be nearly canonical,
\textit{i.e.}, ${\cal K} \simeq \frac{1}{2}(S+S^*)^2
+\frac{1}{2}(T+T^*)^2$, however our results are not drastically changed
for a more generic K\"{a}hler potential provided the $S-T$ mixing is small. Note that unlike the
non-supersymmetric model~\cite{Espinosa:2015eda} there are no radiative
corrections to the amplitude ${\cal A}$. This is due to the
nonrenormalization of the superpotential mass parameter $m_N$.
The nonrenormalization of the superpotential can also suppress the
generation of other potentially dangerous terms like $\cos^2
(\phi/f_\phi)$; since such terms can only come from higher-dimensional
K\"{a}hler type operators like $\int d^4\theta NN\bar{N}\bar{N}$, their
effects should be suppressed by at least the inverse-square of the
mass scale at which the operators are generated, as well as by a loop
factor and a chiral-symmetry violating parameter like $m_N$. Hence,
our model can avoid dangerous radiative corrections, contrary to the
non-supersymmetric model.

\subsection{Conditions on the $\phi, \sigma$ evolution}
\label{sec:scalarconditions}

Consider the cosmological evolution of $\phi$ and $\sigma$ which is driven by the potential (\ref{eq:relaxionpotential}) during an inflationary phase of the Universe. Initially these fields are assumed to have very large positive values,
$\phi, \sigma \gg f_{\phi}$, so that electroweak symmetry is preserved, $H_u =H_d = 0$. For definiteness, we also assume $\overline{m}_N, g_S>0$, while $g_T<0$ with $-g_T \sigma > g_S \phi \gg \overline{m}_N$ so that ${\cal A} > 0$. Of course, there are several other possible combinations of signs which result in a similar evolution of $\phi$ and $\sigma$. The sign of $\lambda$ will be determined below.

Initially, the relaxion field $\phi$ is trapped in a local minimum with a large constant field value because the last term in Eq.~\eqref{eq:relaxionpotential} dominates the potential. This occurs when
\begin{equation}
 |m_S|^2 \phi \ll \frac{\Lambda_N^3}{f_\phi} |{\cal A}(\phi, \sigma,  0)| ~.
\label{eq:phistop}
\end{equation}
As long as this condition is satisfied, $\phi$ remains fixed. However, $\sigma$ is free to roll and therefore during the $\sigma$ evolution a cancellation between the $\phi$ and $\sigma$ terms in the amplitude ${\cal A}$ can remove the barrier and allow $\phi$ to roll. Just prior to $\phi$ rolling, the amplitude is roughly given by ${\cal A} \sim -g_T \sigma \sim g_S \phi$, and thus the above condition becomes
\begin{equation}
|m_S|^2 \ll g_S\frac{\Lambda_N^3}{f_\phi} ~.
\label{eq:phistop2}
\end{equation}

Since there is no potential barrier for $\sigma$, it slow-rolls until it reaches its potential minimum $(\sigma=0)$. Its time-dependence is therefore determined by the following equation:
\begin{equation}
 \frac{d \sigma}{d t} = -\frac{1}{3H_I} \frac{\partial
 V_{\phi,\sigma}}{\partial \sigma}
=-\frac{1}{3H_I}
\left[|m_T|^2 \sigma -\frac{g_T}{\sqrt{2}} \Lambda_N^3 \cos
\left(\frac{\phi}{f_\phi}\right)\right] ~,
\label{eq:sigmaev}
\end{equation}
where $H_I$ is the Hubble parameter during inflation, which is roughly constant. Note that we must assume
\begin{equation}
 |m_T|^2 \sigma \gg |g_T| \Lambda_N^3 ~,
\end{equation}
throughout the relaxation process so that the right-hand side of Eq.~\eqref{eq:sigmaev} is negative-definite and $\sigma$ monotonically decreases. Around the time electroweak symmetry breaking occurs, we have
$g_S \phi_* \sim - g_T \sigma_*$, as we will see below. Combining this relation with  Eq.~\eqref{eq:msusy}, we can
rewrite the above condition as
\begin{equation}
 \frac{g_T^2}{g_S} \ll \frac{m_{\text{SUSY}} f_\phi}{\Lambda_N^3}
\frac{|m_T|^2}{|m_S|} ~.
\label{eq:sigmadecmon}
\end{equation}

As $\sigma$ decreases, the amplitude ${\cal A}$ gets small. Eventually, the condition \eqref{eq:phistop} is violated and $\phi$ starts to evolve. For the relaxion, $\phi$ to roll, the condition
\begin{equation}
 \frac{\Lambda_N^3}{f_\phi}\left\vert{\cal A}(\phi, \sigma, 0)\right\vert \lesssim  |m_S|^2 \phi~,
\label{eq:condphievolv}
\end{equation}
should hold. Using (\ref{eq:simpamp}) this becomes
\begin{equation}
    \left(\frac{g_S}{\sqrt{2}}-\frac{f_\phi |m_S|^2}{\Lambda_N^3}\right) \phi
    \lesssim \overline{m}_N - \frac{g_T}{\sqrt{2}} \sigma\lesssim \left(\frac{g_S}{\sqrt{2}}+\frac{f_\phi |m_S|^2}{\Lambda_N^3}\right)\phi ~,
\label{eq:ineqphislide}
\end{equation}
which must be satisfied until electroweak symmetry breaking occurs.

In addition, $\phi$ must move fast enough once Eq.~\eqref{eq:ineqphislide} is satisfied so that it can track the $\sigma$ evolution. If $\phi$ moves too slowly, ${\cal A}$ will eventually become large and negative, violating the first condition in Eq.~\eqref{eq:ineqphislide}. This will further slow the evolution of $\phi$ until it is eventually trapped
in a minimum with a value that is too large for electroweak symmetry breaking to occur.  Therefore, we require
\begin{equation}
  \left(\frac{g_S}{\sqrt{2}}-\frac{f_\phi |m_S|^2}{\Lambda_N^3}\right)
  \frac{d\phi}{dt} <  - \frac{g_T}{\sqrt{2}} \frac{d\sigma}{dt}~.
\end{equation}
There is no upper bound on the velocity of $\phi$ since this leads to a larger ${\cal A}$ which would in turn slow
its rolling so that $\sigma$ could catch up. During this period of evolution where $\phi$ tracks $\sigma$, the
velocity of $\phi$ is determined by the usual slow-roll condition, as was done for $\sigma$ in Eq.~\eqref{eq:sigmaev}. Thus, the above inequality can be rewritten as follows:
\begin{equation}
 g_S |m_S|^2 \phi > -g_T |m_T|^2 \sigma ~,
\label{eq:condphifollowsig}
\end{equation}
where we have used the condition in Eq.~\eqref{eq:phistop2} and $\frac{d\phi}{d t} =-\frac{m_S^2\phi}{3H_I}$. In particular, when $g_S \phi \sim - g_T \sigma$,
which follows from Eq.~\eqref{eq:ineqphislide} with $\overline{m}_N \ll g_S \phi, |g_T \sigma|$, the above relation reduces to
\begin{equation}
 |m_S| > |m_T| ~.
\label{eq:msgtmt}
\end{equation}
This constraint is consistent with our desired need to suppress the $\sigma$ coupling to MSSM fields which was discussed in Section~\ref{sec:susybreaking}, and results in a Higgs soft mass that predominantly depends on $\phi$.

As long as the above conditions are satisfied, the evolution of $\phi$ is determined by that of $\sigma$, in order to maintain ${\cal A}\simeq 0$. During the rolling of $\phi$, the value of ${\cal D}(\phi)$ in Eq.~\eqref{eq:dphidef} changes unimpeded until it becomes zero for $\phi = \phi_*$. As $\phi$ continues to evolve past this point electroweak symmetry is spontaneously broken. The Higgs VEV will then begin to grow and the amplitude, ${\cal A}$, changes due to the non-zero Higgs VEV, by an amount
\begin{equation}
\Delta {\cal A} = \frac{\lambda}{4} \frac{v^2(\phi)}{M_L} \sin 2\beta~,
\label{eq:dela}
\end{equation}
where $v(\phi)$ is the Higgs VEV given in Eq.~\eqref{eq:higgsvev}. This contribution causes the evolution of $\phi$ to stop tracking the $\sigma$ evolution, with the relaxion field $\phi$ eventually becoming trapped in a local minimum of the periodic potential. As $\sigma$ continues to roll, the amplitude ${\cal A}$ becomes larger and larger reaching a maximum value, when $\sigma$ reaches the minimum of its potential, \textit{i.e.}, $\sigma = 0$.

Let us determine the condition for $\phi$ to stop tracking $\sigma$ after electroweak symmetry breaking. When $\phi < \phi_*$ the condition in Eq.~\eqref{eq:condphievolv}, due to $\phi$ being trapped in a local minimum of the potential, can be written
\begin{equation}
 \frac{\Lambda_N^3}{f_\phi}\left\vert{\cal A}\left(\phi, \sigma,
 \frac{v^2(\phi)}{4}\sin 2\beta\right)\right\vert \lesssim  |m_S|^2 \phi ~,
\label{eq:condphievolvewsb}
\end{equation}
which becomes
\begin{equation}
\scalebox{0.9}{$\displaystyle
 \left(\frac{g_S}{\sqrt{2}}-\frac{f_\phi |m_S|^2}{\Lambda_N^3}\right)
  \phi
+\frac{\lambda \sin(2\beta) {\cal D}(\phi)}{4M_L \lambda_h m_H^2}
\lesssim \overline{m}_N - \frac{g_T}{\sqrt{2}} \sigma
\lesssim
 \left(\frac{g_S}{\sqrt{2}}+\frac{f_\phi |m_S|^2}{\Lambda_N^3}\right)
  \phi
+\frac{\lambda \sin(2\beta) {\cal D}(\phi)}{4M_L \lambda_h m_H^2}
~.$}
\label{eq:ineqphislideew}
\end{equation}
If we now require that this inequality is violated shortly after electroweak symmetry is broken so that $\phi$ stops when $v(\phi)\simeq 246$ GeV, then the left-hand side of Eq.~\eqref{eq:ineqphislideew} should change much
slower than the term with $\sigma$, \textit{i.e.},
\begin{equation}
\left[ \frac{g_S}{\sqrt{2}}-\frac{f_\phi |m_S|^2}{\Lambda_N^3}
+\frac{\lambda \sin 2\beta}{4M_L \lambda_h m_H^2}
\frac{d {\cal D}(\phi)}{d \phi}
\right] \frac{d\phi}{d t} >-\frac{g_T}{\sqrt{2}} \frac{d\sigma}{dt} ~.
\end{equation}
Using the slow-roll condition for $\sigma$ and $\phi$ and the
relation \eqref{eq:phistop2}, this condition becomes
\begin{equation}
\frac{\lambda \sin 2\beta}{4M_L \lambda_h m_H^2}
\phi\frac{d {\cal D}(\phi)}{d \phi}
<
-\frac{1}{\sqrt{2}|m_S|^2}\left(
g_S |m_S|^2 \phi + g_T |m_T|^2 \sigma
\right)
 ~.
\label{eq:condphistopfol}
\end{equation}
From Eq.~\eqref{eq:condphifollowsig}, we see that at least in the vicinity of $\phi = \phi_*$, the right-hand side of
Eq.~\eqref{eq:condphistopfol} is negative (since we have required that Eq.~\eqref{eq:condphifollowsig} is valid
up to $\phi = \phi_*$). As $d{\cal D}(\phi_*)/d\phi_* > 0$, we thus find that $\lambda$ should be negative.

\subsection{Resolving the little hierarchy problem}
\label{sec:littlehierarchy}

In order for the above mechanism to actually resolve the little hierarchy problem, the following conditions should
also be satisfied. First of all, the relaxion field $\phi$ should be trapped by the barrier generated by electroweak symmetry breaking. This requires
\begin{equation}
 |m_S|^2 \phi_* \simeq |\Delta {\cal A}| \frac{\Lambda^3_N}{f_\phi} ~.
\end{equation}
By using Eqs.~\eqref{eq:dela} and \eqref{eq:msusy}, we obtain
\begin{equation}
 |m_S| \simeq \frac{|\lambda| \sin 2\beta}{4 M_L}
\frac{v^2 \Lambda_N^3}{m_{\text{SUSY}}f_\phi^2}~,
\label{eq:lamovlam}
\end{equation}
where $v\simeq 246$~GeV. By substituting this equation into
Eq.~\eqref{eq:condphistopfol}, and using
\begin{equation}
 m_H \sim m_{\text{SUSY}}~, ~~~~ \phi\frac{d {\cal D}(\phi)}{d \phi}
\sim m_{\text{SUSY}}^4~,~~~~ g_S \phi_* \sim -g_T \sigma_* ~,
\end{equation}
plus the fact that $m_h^2=\lambda_hv^2$, we obtain the constraint
\begin{equation}
 g_S \frac{m_h^2}{m^2_{\text{SUSY}}}  \frac{\Lambda_N^3}{f_\phi}
\lesssim \frac{|m_S|^2}{1-\frac{|m_T|^2}{|m_S|^2}} ~,
\label{eq:phidecop}
\end{equation}
where $m_h\simeq 125$~GeV is the mass of the SM-like Higgs boson.

Second, the evolution of $\phi$ and $\sigma$ should satisfy the slow-roll conditions. Given the constraint \eqref{eq:msgtmt}, if $\phi$ satisfies the slow-roll condition, then so does $\sigma$. Thus, it is sufficient
to just consider the slow roll condition of $\phi$, where the second time derivative
$\frac{d^2\phi}{dt^2}$ should be much smaller than the Hubble friction term $|3H_I\frac{d\phi}{dt}|$
and the gradient of the potential energy $\frac{\partial V_{\phi,\sigma}}{\partial\phi}$. This leads to the
constraint
\begin{equation}
 |m_S| \ll H_I ~.
\label{eq:slowrol}
\end{equation}

\subsubsection{Constraints from inflationary dynamics}

In the relaxion mechanism, inflation is driven by a separate inflaton sector that is not explicitly given in our model. This sector should dominate the vacuum energy and not receive a back-reaction from the relaxion sector. In particular, the potential energy carried by $\phi$ and $\sigma$ should be much smaller than the inflaton energy:
\begin{equation}
 \frac{1}{2}|m_S|^2 \phi^2 ~, ~ \frac{1}{2}|m_T|^2 \sigma^2
\ll 3 H_I^2 M_P^2 ~,
\label{eq:infencond}
\end{equation}
which gives a lower bound on the Hubble parameter during inflation:
\begin{equation}
 H_I \gg m_{\text{SUSY}}\frac{f_\phi }{M_P} ~.
\label{eq:infen}
\end{equation}

The energy of inflation breaks SUSY, and its effects can be included into the $F$-term of the
inflaton chiral multiplet: $\Phi_I = \dots + H_IM_P \theta\theta$. This generically gives a gravity-mediated contribution to the soft masses of order
\begin{equation}
 \delta m_{\text{SUSY}} \sim H_I~.
\end{equation}
In order that this contribution does not affect the scale of electroweak symmetry breaking, we require
\begin{equation}
 H_I \lesssim v ~,
\label{eq:infsoft}
\end{equation}
so that when inflation ends, the Higgs VEV does not change
significantly.\footnote{This bound assumes $F$-term inflation, whereas
models of $D$-term inflation may be able to avoid this bound
\cite{Binetruy:1996xj}.} As long as $\Lambda_N \gg v$, this condition
also means $H_I \ll \Lambda_N$, which ensures that a periodic potential
is formed during inflation.

In order for the relaxion mechanism to work, the evolution of $\phi$ and $\sigma$ should be dominated by classical rolling, \textit{i.e.}, the effects of quantum fluctuations on the evolution of these fields
should be negligible. During a period of one Hubble time $H_I^{-1}$, the field $\sigma$ changes by
\begin{equation}
 \left\vert\frac{d\sigma}{d t} H_I^{-1}\right\vert \sim
\frac{|m_T|^2 \sigma}{3 H_I^2}
>\frac{|m_T|^2 \sigma_*}{3 H_I^2}
\sim \frac{|m_T|^2 g_S\phi_*}{3 |g_T| H_I^2} ~.
\end{equation}
This should be larger than the typical size of quantum fluctuations of ${\cal O}(H_I)$, which gives an upper bound on $H_I$:
\begin{equation}
 H_I^3 \ll  \frac{g_S}{|g_T|} |m_T|^2 \phi_*
\sim
 \frac{g_S |m_T|^2 f_\phi m_{\text{SUSY}}}{|g_T| |m_S|} ~.
\label{eq:classicalrol}
\end{equation}

Finally, let us estimate the number of $e$-folds required in this
setup. For the relaxion field $\phi$ to naturally scan the critical
value $\phi_*$, its excursion range during inflation, $\Delta \phi$,
should be larger than $\phi_*$:
\begin{equation}
 \Delta \phi > \phi_* \sim \frac{f_\phi m_{\text{SUSY}}}{|m_S|} ~.
 \label{delphi}
\end{equation}
The number of $e$-folds $N_e$ is then given by
\begin{equation}
 N_e \simeq \frac{H_I \Delta \phi}{\left\vert
	\frac{d\phi}{dt}\right\vert}
\simeq
\frac{3 H_I^2 \Delta \phi}{\left\vert
 \frac{\partial V}{\partial \phi}\right\vert}
\gtrsim \frac{H_I^2}{|m_S|^2}
=10^{14} \times \left(\frac{H_I}{1~{\rm GeV}}\right)^2
\left(\frac{10^{-7}~{\rm GeV}}{|m_S|}\right)^2
~.
\label{eq:nefolds}
\end{equation}
This is a large number, although much smaller than those predicted in
Refs.~\cite{Espinosa:2015eda, Batell:2015fma}. Note also that (\ref{delphi}) implies that $\phi$ can have super-Planckian field values for large $m_{\text{SUSY}}$ and small $m_S$.  These field values can be explained by the UV description given in Appendix~\ref{sec:uvcomp}.

\begin{figure}[t]
\centering
\includegraphics[clip, width = 0.7 \textwidth]{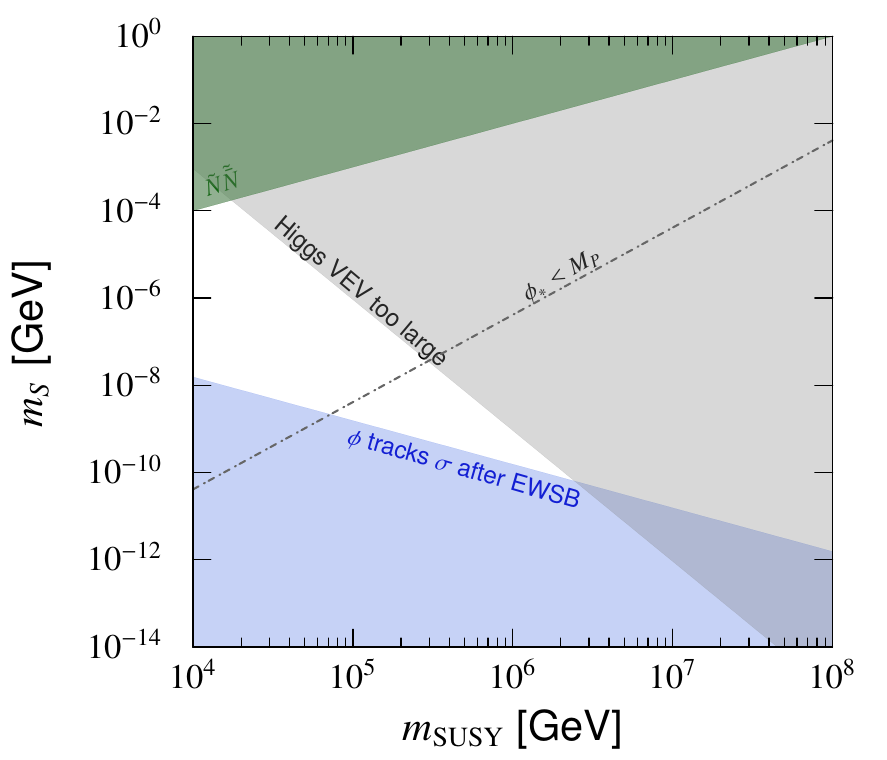}
\caption{{The allowed parameter region in the $m_{\rm SUSY}$--$m_S$ plane, where $\zeta = 10^{-8}$, $r_{TS}=0.1$, $r_\Lambda = 1$, and $r_{\rm SUSY} = 1$.}}
\label{fig8100}
\end{figure}

\begin{figure}[t]
\centering
\includegraphics[clip, width = 0.7 \textwidth]{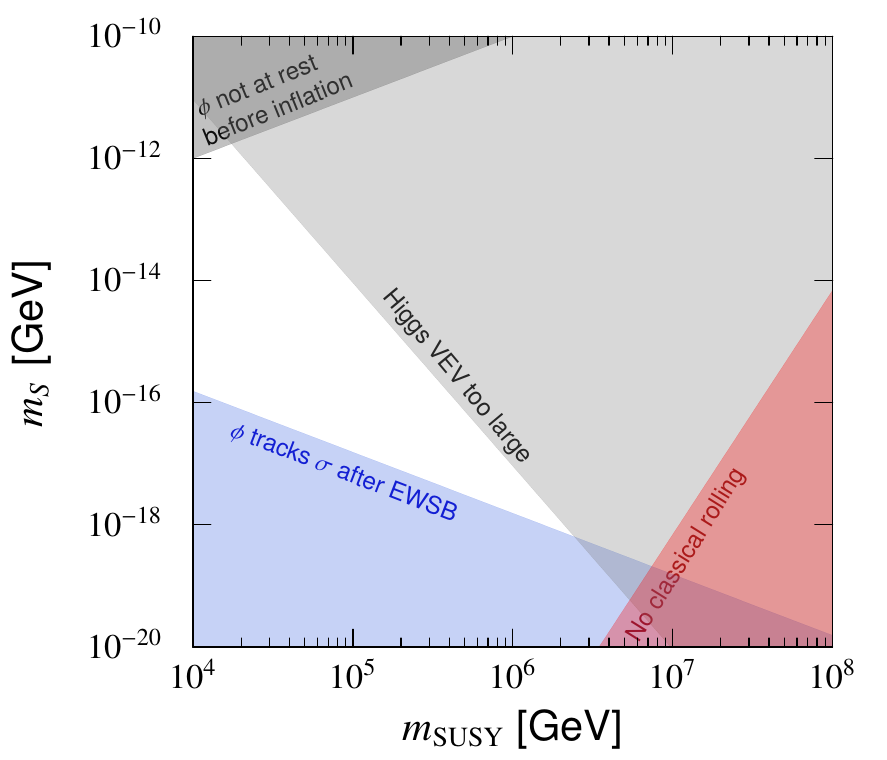}
\caption{{The allowed parameter region in the $m_{\rm SUSY}$--$m_S$ plane, where $\zeta = 10^{-8}$, $r_{TS}=0.1$, $r_\Lambda = 10^{-4}$, and $r_{\rm SUSY} = 10^{-4}$.}}
\label{fig8144}
\end{figure}

\begin{figure}[t]
\centering
\includegraphics[clip, width = 0.7 \textwidth]{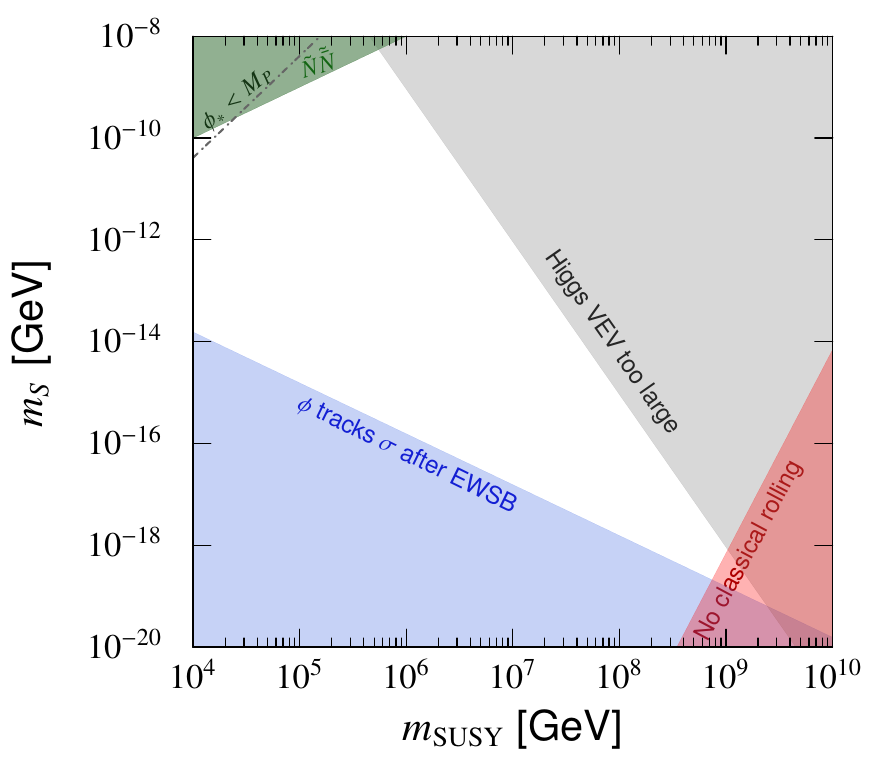}
\caption{{The allowed parameter region in the $m_{\rm SUSY}$--$m_S$ plane, where $\zeta = 10^{-14}$, $r_{TS}=0.1$, $r_\Lambda = 1$, and $r_{\rm SUSY} = 1$.}}
\label{fig14100}
\end{figure}

\subsubsection{Summary of Constraints}
\label{sec:summaryplots}

The various constraints on the two-field supersymmetric model can be shown graphically to reveal the allowed regions of parameter space. In order to reduce the number of independent parameters we assume the following
\begin{align}
 g_S &= \zeta \frac{m_S}{f_\phi}, ~~~~
 g_T = \zeta \frac{m_T}{f_\sigma}, ~~~~
 f\equiv f_\phi = f_\sigma, ~~~~ \nonumber \\
 r_{TS} &\equiv \frac{m_T}{m_S} ,~~~~
 r_\Lambda \equiv \frac{\Lambda_N}{f}, ~~~~
 r_{\text{SUSY}} \equiv \frac{m_{\text{SUSY}}}{f},
~~~ M_L = m_{\rm SUSY}~,
\end{align}
where $\zeta$ is a dimensionless parameter. The constraints are plotted in the $m_{\rm SUSY}$--$m_S$ plane for fixed values of $\zeta$, $r_{TS}$, $r_\Lambda$, and $r_{\rm SUSY}$.

{In Figure~\ref{fig8100}, we show the allowed parameter region for $\zeta
= 10^{-8}$, $r_{TS}=0.1$, $r_\Lambda = 1$, and $r_{\rm SUSY} =1$.
The light gray region is excluded since $\phi$ cannot be
stopped by the periodic potential after electroweak symmetry breaking
with $|\lambda|\leq \frac{M_L v^2}{\Lambda_N^3}$, which follows from the
requirement that the contribution of the higher dimensional operator in
Eq.~\eqref{eq:wn} to the $B\mu$ term should be smaller than $v^2$. } 
In the blue-shaded region, the two-field relaxion mechanism does not work since 
$\phi$ keeps tracking $\sigma$ even after electroweak symmetry breaking (Eq.~\eqref{eq:phidecop}). 
The dash-dotted line corresponds to $\phi_*=M_P$,
which represents the boundary between sub-Planckian and super-Planckian
field values of $\phi$ (see Eq.~\eqref{eq:phivalue}).
{Finally, the green region is disfavored as the potential may become
unstable in the direction of $\widetilde{N} \widetilde{\bar{N}}$.
Note that the maximum allowed value of $m_{\text{SUSY}}$ is $\sim 2\times 10^6$ GeV, 
corresponding to $m_S \sim 10^{-10}\, \text{GeV}$.}

{The parameter values in Figure~\ref{fig8144} are the same as in Figure~\ref{fig8100} except that now 
$r_\Lambda= r_{\rm SUSY}= 10^{-4}$. 
In particular this reduces the allowed values of $m_S$ and the dark gray region is excluded since initially $\phi$ is not trapped at a potential minimum due to an insufficient barrier height (Eq.~\eqref{eq:phistop2}). }
Notice that in this case $m_{\rm SUSY} ={\cal O}(1)$~PeV corresponds to $f_\phi = {\cal O}(10^{10})$~GeV, which
is right in the middle of the usual axion window. This may allow the possibility
to construct a model where both $S$ and the QCD axion are generated by the
same dynamics at this energy scale.
{In Figure~\ref{fig14100} we have instead changed the value of $\zeta$
to be $\zeta = 10^{-14}$ as compared to Figure~\ref{fig8100}, which
effectively reduces the values of $g_S$ and $g_T$. In this case the
region for the allowed values of $m_S$ decreases since the constraint
from $\phi$ not tracking $\sigma$ (blue region) is not as severe and the
maximum allowed value of $m_{\rm SUSY}$ increases to $\sim 10^9$
GeV. The upper bound on $m_{\rm SUSY}$ now arises from the fact that the
classical rolling condition \eqref{eq:classicalrol} is inconsistent with
the condition that inflation dominates the vacuum energy---the
corresponding excluded area is illustrated by the red shaded region. Note that in all figures there is always an allowed region for $m_{\rm SUSY}$ that includes the PeV scale ($10^6$ GeV), which can be obtained for $10^{-17} ~{\rm GeV} \lesssim m_S \lesssim 10^{-9}$~GeV depending on the range of the other parameters considered in the figures.}

Finally, in the allowed region, we can always find a value of $H_I$ which satisfies the
constraints \eqref{eq:slowrol}, \eqref{eq:infen}, \eqref{eq:infsoft}, and
\eqref{eq:classicalrol}. The lower bound on $H_I$ is given by
\begin{equation}
 H_I > \text{max}\left\{|m_S|,~4\times 10^{-9}~\text{GeV}\times
\left(\frac{m_{\rm SUSY}}{10^5~{\rm GeV}}\right)^2
\left(\frac{1}{r_{\rm SUSY}}\right)
 \right\}~,
\end{equation}
while the upper bound is
\begin{equation}
 H_I < \text{min}\left\{v,~4.6~{\rm GeV}\times
\left(\frac{r_{TS}}{0.1}\right)^{\frac{1}{3}}
\left(\frac{1}{r_{\rm SUSY}}\right)^{\frac{1}{3}}
\left(\frac{|m_S|}{10^{-7}~{\rm GeV}}\right)^{\frac{1}{3}}
\left(\frac{m_{\rm SUSY}}{10^5~{\rm GeV}}\right)^{\frac{2}{3}}
\right\}~.
\end{equation}
Therefore, $H_I$ can be larger than that predicted in
ref.~\cite{Batell:2015fma}, where $H_I$ had to be smaller than the QCD
scale $\sim 300$~MeV.

\section{Conclusion}
\label{sec:conclusion}

We have presented a supersymmetric relaxion mechanism that can provide a solution to the little hierarchy problem in supersymmetric models. Our supersymmetric extension is based upon the nonsupersymmetric two-field relaxion model of Ref.~\cite{Espinosa:2015eda}. Since the two-field model assumes no new source of electroweak symmetry breaking beyond that due to the Higgs field, the scale of strong dynamics can be arbitrarily large. This preserves the QCD axion solution to the strong CP problem and allows the cutoff scale to be significantly increased beyond the TeV scale.  However, the periodic potential of the relaxion,
$\phi$, has a large amplitude that must now be neutralized by a second field, $\sigma$ which has no couplings to the Higgs sector.

In the supersymmetric extension, the scalar fields $\phi$, $\sigma$ are
identified with the imaginary scalar field components of two chiral
superfields $S$, $T$ respectively, which transform under shift symmetries each associated with a global symmetry. In particular the shift symmetry associated with $T$ prevents a $\sigma$-Higgs coupling at the renormalizable level, realizing a crucial feature of the nonsupersymmetric model. A nontrivial relaxion potential is generated when the shift symmetries are explicitly broken and cause the two scalar fields to dynamically evolve from their initially large field values. This induces large $F$-terms which break supersymmetry and therefore as the relaxion rolls, it scans the soft supersymmetric masses. Electroweak symmetry is broken when the soft supersymmetric mass scale is of order the $\mu$-term. In this way, an apparently tuned cancellation can be explained as the result of a dynamical mechanism~\cite{Batell:2015fma}.

The conditions for cosmological evolution of the scalar fields and the requirements of obtaining a correct electroweak Higgs VEV, restrict the parameter space of our supersymmetric two-field relaxion model.
In particular the Hubble scale during inflation must at least satisfy
$H_I\lesssim v$, where $v$ is the electroweak VEV, a condition necessary
to prevent shifts from the inflation sector, although depending on parameter values 
the classical rolling condition can give a more stringent upper bound on
$H_I$. The soft supersymmetric mass scale plays the role of the cutoff and for
typical values of the allowed couplings we find that this scale can be
as large as $10^9$ GeV for a range
{$10^{-18}\, {\rm GeV} \lesssim m_S \lesssim 10^{-4}\, {\rm GeV}$, }
where $m_S$ is an explicit shift-symmetry breaking parameter.  When the mediation scale is of order the global symmetry breaking scale the soft masses can be of order the PeV ($10^6$ GeV) scale, while gaugino masses are suppressed by a loop factor. This is similar to a split-SUSY spectrum, except for $A$-terms which are not loop suppressed as occurs in models based on anomaly mediation. There is also the possibility for extra suppression if the matter and Higgs fields do not transform under the shift symmetry. Furthermore, when the mediation scale is much larger than the global symmetry breaking scale, the soft supersymmetric masses are induced radiatively from gaugino masses, giving rise to a gaugino-mediation type spectrum. With further modifications one can obtain soft masses to be near the TeV scale. Thus, we see that the tuning is avoided in supersymmetric models, with a more detailed survey of the sparticle spectrum left for future work.

The larger values of $m_S$ allow some of the cosmological constraints to be weakened compared
to the previous implementations of the relaxion mechanism. For instance we find that the minimum
number of $e$-folds required for the relaxion to complete its cosmological evolution can be substantially smaller, $\gtrsim 10^{14}$ for $m_S\sim 10^{-7}$ GeV and $H_I\sim 1$~GeV. In fact for larger values of
$m_S$ there are regions of the parameter space for which it is possible to have sub-Planckian values during the evolution of the relaxion. This obviates the need to explain how super-Planckian
field excursions can arise in the effective field theory.

In the case when the scalar fields $\phi$, $\sigma$ have large field values, supergravity effects can lead to potentially large contributions in the relaxion potential. To preserve the results in the global supersymmetry limit, these supergravity effects need to be cancelled by an additional source of supersymmetry breaking with an almost no-scale structure. Generically this causes the gravitino to be much heavier than the relaxino $\widetilde\phi$, and therefore the relaxino is typically the lightest supersymmetric particle. Provided the relaxino is heavy enough ($\gtrsim \text{keV}$), this allows for the possibility that the relaxino is the dark matter candidate. Thus, together with the axion-like particles $\phi$ and $\sigma$, these are the only new states at low energies.

Finally, our supersymmetric model is an effective description at low energies that in some cases requires super-Planckian field values of the scalar fields. This is potentially inconsistent with an axion interpretation, but using recent ideas of~Ref.~\cite{Kaplan:2015fuy,Choi:2015fiu}, a large number of axion fields can be used to provide a UV description of our model. In particular the explicit shift-symmetry breaking parameters ($m_{S,T}, g_{S,T})$ can be derived from an axion-like potential in the UV model. Thus, the fact that a consistent field theory description exists for our effective model enhances the relaxion mechanism as a possible solution for restoring naturalness in supersymmetric models.

\section*{Acknowledgments}

We thank Sang Hui Im for several useful discussions about this work.
The work of T.G. and N.N. is supported by the DOE grant DE-SC0011842 at the
University of Minnesota, and Z.T. is supported by the University of
Minnesota.

\section*{Appendix}
\appendix

\section{Explicit formulae for the two-field supersymmetric relaxion model}
\label{sec:formulae}
\renewcommand{\theequation}{A.\arabic{equation}}
\setcounter{equation}{0}

In this section, we summarize the supersymmetric Lagrangian and present explicit formulae for the
supersymmetry-breaking parameters.

\subsection{Lagrangian}
\label{sec:lagrangian}

First, we summarize the Lagrangian of our model. The most generic K\"{a}hler potential 
invariant under the transformations ${\cal S}_S$, (\ref{Strans}) and ${\cal S}_T$, (\ref{Ttrans}) 
is given as follows:
\begin{align}
 K &= \kappa (S + S^*, T + T^*)
+Z_i (S + S^*, T + T^*) \Phi_i^* e^{2V} \Phi_i \nonumber \\
&+ \left[
U(S + S^*, T + T^*) e^{-\frac{q_HS}{f_\phi}}H_u H_d
+\text{h.c.}
\right]~,
\label{eq:kahler}
\end{align}
where $\Phi_i = Q_i, H_u, H_d, N, \bar{N}$. The superpotential is, on the
other hand, given by
\begin{equation}
 W = W_{\text{gauge}} + W_{\text{Yukawa}}+ W_{\mu} + W_{S, T}
+ W_N~,
\label{eq:superpotential}
\end{equation}
where
\begin{align}
 W_{\text{gauge}} &= \left(\frac{1}{2g_a^2} - i\frac{\Theta_a}{16\pi^2}
-\frac{c_aS}{16\pi^2 f_\phi}\right){\rm Tr}{\cal W}_a{\cal W}_a
~,\\[3pt]
 W_{\rm Yukawa} &= y_u Q \overline{U}H_u + y_d Q\overline{D} H_d
+y_e L \overline{E} H_d ~,\\
 W_{\mu} &= \mu_0 e^{-\frac{q_HS}{f_\phi}} H_u H_d ~,\\
 W_{S, T} &= \frac{1}{2}m_SS^2 +\frac{1}{2}m_TT^2 ~,\label{eq:wst2}\\
 W_N &= m_N N \bar{N} +
ig_S S N \bar{N} +ig_T T N \bar{N} +
 \frac{\lambda}{M_L} H_u H_d N \bar{N}~.
\label{eq:wn2}
\end{align}
The index $a$ sums over the appropriate groups of ${\rm SM} \times {\rm
SU}(N)$, $g_a$ and $\Theta_a$ are the corresponding gauge coupling
constants and the vacuum angle, respectively, and $c_a$ is a constant
which depends on the UV completion of this model. The higher-dimensional 
operator in (\ref{eq:wn2}) is assumed to be generated at a UV scale, $M_L$.

\subsection{Scalar potential}

Next, we calculate the scalar potential in our model from the K\"{a}hler
potential (\ref{eq:kahler}) and the superpotential (\ref{eq:superpotential}).
The Lagrangian for $S$ and $T$ without derivatives is given by
\begin{align}
 {\cal L}&= \bm{F}^\dagger {\cal K}({s}, {\tau}) \bm{F}
+\left(\bm{m}\cdot \bm{F}  +i\bm{g}\cdot \bm{F} \widetilde{N}
\widetilde{\bar{N}}+ {\rm h.c.}\right)~,
\label{eq:noderiv}
\end{align}
where 
\begin{equation}
 {\cal K} =\frac{1}{2}
\begin{pmatrix}
 \frac{\partial^2 \kappa}{\partial {s}^2} &
 \frac{\partial^2 \kappa}{\partial {s} \partial {\tau}} \\
 \frac{\partial^2 \kappa}{\partial {s} \partial {\tau}} &
 \frac{\partial^2 \kappa}{\partial {\tau}^2}
\end{pmatrix}
,
~~~
\bm{F} = 
\begin{pmatrix}
 F_S \\ F_T
\end{pmatrix}
, ~~~
\bm{g} = 
\begin{pmatrix}
 g_S \\ g_T
\end{pmatrix}
,~~~
\bm{m} = \frac{1}{\sqrt{2}}
\begin{pmatrix}
 m_S (s+i\phi) \\ m_T(\tau +i\sigma)
\end{pmatrix}
~.
\end{equation}
Note that we have set the MSSM scalar fields including $H_u$ and $H_d$ to zero to simplify the 
expression for the scalar potential. Generically, ${\rm det}\,{\cal K} \neq 0$, and thus we can solve the
equation of motion for $\bm{F}$ as
\begin{equation}
 \bm{F} = -{\cal K}^{-1} \left(\bm{m}+i\bm{g}\widetilde{N}\widetilde{\bar{N}}\right)^*~.
\end{equation}
By substituting this expression into (\ref{eq:noderiv}), we obtain the potential term containing the relaxion field:
\begin{equation}
 V_{S, T} = \left(\bm{m}+i\bm{g}\widetilde{N}
\widetilde{\bar{N}}\right)^\dagger {\cal K}^{-1} 
\left(\bm{m}+i\bm{g}\widetilde{N} \widetilde{\bar{N}}\right) ~.
\end{equation}
The $F$-terms of $N$ and $\bar{N}$ also contribute to the scalar potential which evaluates as
\begin{equation}
 V_N^F = \left\vert m_N + ig_S\left(\frac{s+ i\phi}{\sqrt{2}}\right) 
+i g_T\left(\frac{\tau+ i\sigma}{\sqrt{2}}\right)\right\vert^2
\left(\frac{1}{Z_{\bar{N}}}|\widetilde{{N}}|^2
+\frac{1}{Z_N}|\widetilde{\bar{N}}|^2
\right)~.
\end{equation}
In addition, $\widetilde{{N}}$ and $\widetilde{\bar{N}}$ have the
$D$-term contribution coming from the SU($N$) gauge interaction:
\begin{equation}
 V^D_N = \frac{g_N^2}{2}\left(\widetilde{N}^* T^A \widetilde{N} +
\widetilde{\bar{N}} T^A \widetilde{\bar{N}}^* \right)^2 ~,
\end{equation}
where $T^A$ denotes the SU($N$) generators. These two contributions
allow $\widetilde{N}$ and $\widetilde{\bar{N}}$ to remain at the
origin, $\widetilde{N} = \widetilde{\bar{N}} = 0$, as we discuss in Section \ref{sec:relaxion}. 
Therefore, to study the potential structure relevant to the relaxion mechanism, it is
sufficient to consider the following simplified potential, which is obtained by
setting $\widetilde{N} = \widetilde{\bar{N}} = 0$ in $V_{S, T}$: 
\begin{equation}
 V =  \bm{m}^\dagger {\cal K}^{-1} 
\bm{m} ~.
\end{equation}
This potential has a trivial minimum at $\bm{m} = 0$, \textit{i.e.}, $s
= \phi = \tau = \sigma = 0$. At this minimum, $V = 0$, and thus SUSY is
not broken. However the relaxion scenario assumes that at the beginning 
of the cosmological evolution, $\phi$ and $\sigma$ have very large field values, and thus are 
far away from this minimum. This implies that SUSY is broken during the evolution of
$\phi$ and $\sigma$. 

Assuming a fixed value of $\phi$ and $\sigma$, the potential minimum is determined 
by the condition, 
\begin{equation}
 \frac{\partial}{\partial {s}} {\cal K}^{-1} ({s}, {\tau})
\simeq \frac{\partial}{\partial {\tau}} {\cal K}^{-1} ({s},
{\tau}) \simeq 0 ~,
\end{equation}
for $|\phi|, |\sigma| \gg {s}, {\tau}$. If $\kappa$ is a generic
function, then we expect that ${s}/f_\phi$ and
${\tau}/f_\sigma$ are ${\cal O}(1)$ at this point. Note that the
minimum does not depend on $\phi$ and $\sigma$ as long as they have very large values 
\cite{Batell:2015fma}, since the above condition is independent of these values. Hence, we can
think of ${s}$ and ${\tau}$ as constant during the relaxion process.

\subsection{Soft mass parameters}
\label{sec:softmass}

The non-zero field values of $\phi$ and $\sigma$ effectively generate soft
SUSY-breaking mass terms in the MSSM sector. To compute these soft
masses, we first expand the functions $Z_i$, $U$, and $e^{-\frac{q_HS}{f_\phi}}$ in 
terms of the scalar component fields of $S$ and $T$:
\begin{align}
 Z_i(S+S^*, T+T^*)
&= Z_i(\sqrt{2}{s}, \sqrt{2}{\tau})
+\left[\bm{Z}_i \cdot \bm{F} \theta\theta + \text{h.c.}\right]
+\bm{F}^\dagger {\cal Z}_i \bm{F} \theta\theta \overline{\theta}
 \overline{\theta}  ~, \\
 U(S+S^*, T+T^*)
&= \bm{U} \cdot \bm{F}^* \overline{\theta} \overline{\theta}
+\bm{F}^\dagger {\cal U} \bm{F} \theta\theta \overline{\theta}
 \overline{\theta}  ~, \\
e^{-\frac{q_HS}{f_\phi}} &= e^{-\frac{q_H(s+i\phi)}{\sqrt{2}f_\phi}}
\left[1-\frac{q_H}{f_\phi} F_S \theta^2\right] ~,
\end{align}
where
\begin{equation}
 {\bf X} = \frac{1}{\sqrt{2}}
\begin{pmatrix}
 \frac{\partial X}{\partial {s}} \\
 \frac{\partial X}{\partial {\tau}}
\end{pmatrix}
~,~~~~
{\cal X} =\frac{1}{2}
\begin{pmatrix}
  \frac{\partial^2 X}{\partial {s}^2} &
 \frac{\partial^2 X}{\partial {s} \partial {\tau}} \\
 \frac{\partial^2 X}{\partial {s} \partial {\tau}} &
 \frac{\partial^2 X}{\partial {\tau}^2}
\end{pmatrix}
~,
\end{equation}
for $X = Z_i, U$. 

To obtain the physical soft masses, we canonically normalize the fields
$\Phi_i$ by re-scaling them as
\begin{equation}
 \Phi_i \to Z_i^{-\frac{1}{2}}\left[1 - Z_i^{-1} \bm{Z}_i \cdot \bm{F}
\theta\theta \right]\Phi_i ~.
\end{equation}
Then, the soft masses are evaluated as follows:
\begin{align}
 \widetilde{m}^2_i &=  -\frac{1}{Z_i}\bm{F}^\dagger {\cal Z}_i \bm{F}
+\frac{1}{Z_i^2}\left\vert \bm{Z}_i \cdot \bm{F}\right\vert^2 ~, \\
A_{ijk}&= y_{ijk}\left[\frac{\bm{Z}_i}{Z_i} + \frac{\bm{Z}_j}{Z_j} 
+ \frac{\bm{Z}_k}{Z_k} \right]\cdot \bm{F}  ~, \\
\mu &= \frac{1}{(Z_{H_u} Z_{H_d})^{\frac{1}{2}}}
 e^{-\frac{q_H(s+i\phi)}{\sqrt{2}f_\phi}} \left[\mu_0 + \bm{U}\cdot
 \bm{F}^* \right] ~, \\[3pt]
B\mu &= \mu \left[\frac{q_H}{f_\phi} F_S + \left(
\frac{\bm{Z}_{H_u}}{Z_{H_u}}
 + \frac{\bm{Z}_{H_d}}{Z_{H_d}}\right)
 \cdot \bm{F}\right] - \frac{1}{(Z_{H_u} Z_{H_d})^{\frac{1}{2}}}
 e^{-\frac{q_H(s+i\phi)}{\sqrt{2}f_\phi}} \bm{F}^\dagger {\cal U} \bm{F}
 ~, \\
M_a &= \frac{g_a^2}{16\pi^2}\left[
c_a \frac{F_S}{f_\phi} 
-2\sum_{i}\frac{T_{a}^i}{Z_i} \bm{Z}_i \cdot \bm{F} 
\right]
+\frac{g_a^2 B}{8\pi^2} f \left(\frac{\mu^2}{m_H^2}\right)
(\delta_{a1} + \delta_{a2})
\label{eq:gauginomass}
~,
\end{align}
where we have included the wave-function renormalization factors into the
renormalized Yukawa couplings $y_{ijk}$. The last term in
Eq.~\eqref{eq:gauginomass} comes from a gauge-mediation effect from
$H_u$ and $H_d$, where $f(x) = (x\ln x)/(x-1)$ and $m_H$ is the heavy
Higgs mass. These expressions can then be used to obtain the sparticle mass spectrum
discussed in Section \ref{sec:mssm}.

\section{A UV completion of the two-field supersymmetric relaxion model}
\label{sec:uvcomp}
\renewcommand{\theequation}{B.\arabic{equation}}
\setcounter{equation}{0}

The two-field supersymmetric relaxion model is an effective theory which is only
valid below the energy scales $f_\phi$ and $f_\sigma$. Moreover,
the non-compact nature of $\phi$ and $\sigma$ is given explicitly by shift-symmetry
breaking terms whose origin is not specified. However, as discussed in
Ref.~\cite{Gupta:2015uea}, the realization of such small shift-symmetry breaking effects
in any quantum field theory, especially for a field like $\phi$ which has a periodic
potential, is highly non-trivial.

In this Appendix, we present a UV completion of the two-field supersymmetric relaxion
model given in Sec.~\ref{sec:model} which can explain the origin of the small shift-symmetry
breaking effects. We will use the ideas presented in Ref.~\cite{Kaplan:2015fuy} (see also
Refs.~\cite{Choi:2015fiu, Fonseca:2016eoo, Kim:2004rp, Harigaya:2014eta, Choi:2014rja,
Higaki:2014pja,Kappl:2014lra, Ben-Dayan:2014zsa, Dine:2014hwa,
Yonekura:2014oja, Bai:2014coa, Harigaya:2014rga, delaFuente:2014aca, Peloso:2015dsa}).
Specifically we will need to realize the shift-symmetry breaking term and the particular
structure of the periodic potential for the field $S$, which contains the relaxion field $\phi$.
A similar UV completion may be considered for the field $T$, but this is not necessary
because we do not need a periodic potential for $\sigma$.

Consider a set of chiral superfields $\phi_i$, $\bar{\phi}_i$,
$S_i$ $(i = 0, \dots, N)$. These fields are assumed to interact with
each other through the following superpotential:
\begin{equation}
 W_{\rm UV} = \sum_{i=0}^{N} \lambda_i S_i \left(\phi_i \bar{\phi}_i
-f_i^2\right) +
\epsilon \sum_{i=0}^{N-1} \left(\bar{\phi}_i \phi_{i+1}^2
+\phi_i \bar{\phi}^2_{i+1}
\right) ~.
\label{eq:supmulax}
\end{equation}
The first two terms are symmetric under a U(1)$^{N+1}$ global symmetry for which the $S_i$
are neutral, $Q_{S_i} =0$, while $\phi_i$ and $\bar{\phi}_i$ have opposite charges, $Q_{\phi_i} =
-Q_{\bar{\phi}_i}$. The second term explicitly breaks the U(1)$^{N+1}$ symmetry leaving only a single U(1) symmetry.
The parameter $\epsilon$ is assumed to be small, and thus the U(1)$^{N+1}$ global symmetry remains an approximate symmetry. This remaining symmetry corresponds to the charge assignment
\begin{equation}
 Q_{S_i} =0, ~~~~~~Q_{\phi_i} =
-Q_{\bar{\phi}_i} = \frac{1}{2^i} Q_{\phi_0} ~,
\end{equation}
where we normalize the charge assignment to $Q_{\phi_0} = 1$.

The first two terms in Eq.~\eqref{eq:supmulax} lead to spontaneous
symmetry breaking of the approximate global U(1)$^{N+1}$. This breaking gives
rise to $N+1$ approximately massless chiral superfields and $2(N+1)$
massive ones. Below the symmetry breaking scale, we can
parametrize the chiral superfields $\phi_i$ and $\bar{\phi}_i$ as
\begin{equation}
 \phi_i = f_i e^{\frac{\Pi_i}{f_i}} ~, ~~~~~~
 \bar{\phi}_i = f_i e^{-\frac{\Pi_i}{f_i}} ~.
\end{equation}
In this case, the effective superpotential below the symmetry-breaking
scale is given by
\begin{equation}
 W_{\text{eff}} = 2\epsilon \sum_{i=0}^{N-1}
f_i f_{i+1}^2 \cosh \left[
\frac{\Pi_i}{f_i} - \frac{2 \Pi_{i+1}}{f_{i+1}}
\right] ~.
\end{equation}
Among these terms, let us focus on the quadratic terms:
\begin{equation}
 W_{\text{eff}} \supset
\epsilon \sum_{i=0}^{N-1} f_i f_{i+1}^2  \left[
\frac{\Pi_i}{f_i} - \frac{2 \Pi_{i+1}}{f_{i+1}}
\right] ^2
\equiv \sum_{i, j = 0}^{N} \Pi_i {\cal M}_{ij} \Pi_j ~.
\end{equation}
Since there remains a U(1) symmetry, the matrix ${\cal M}$ must have one zero eigenvalue. Up to an overall normalization factor, the corresponding eigenvector is given by
\begin{equation}
 {\cal M}_{ij} z_j = 0 ~, ~~~~~z_i = \frac{f_i}{2^i f_0} ~.
\end{equation}
This massless mode will be identified as the chiral superfield $S$ that contains the relaxion field:
\begin{equation}
 S = c_N \sum_{i=0}^{N} \frac{f_i}{2^i f_0} \Pi_i ~,
\end{equation}
where $c_N$ is a normalization factor. Notice that the $\Pi_N$ component
of $S$ has an exponentially suppressed (a factor of $\sim 2^{-N}$)
coefficient.

To reproduce the model discussed in Sec.~\ref{sec:model} we first identify the remnant U(1) symmetry as the shift symmetry of the $S$ field, ${\cal S}_S$. If the field $\phi_0$ couples to superfields which are chirally charged under the remnant U($1$) and these chiral fields are charged under the SU($N$) gauge symmetry that confines at the scale $\Lambda_N$, then when the chiral superfields are integrated out they generate an axion-like coupling of $S$ with the SU($N$) gauge fields having a decay constant $f_\phi \sim f_0$. This generates the relaxion potential $\propto \cos(\phi_0/f_0)=\cos(\phi/f_\phi)$.

Instead the soft (and very small) ${\cal S}_S$-symmetry breaking term is generated if $\phi_N$ couples to a different set of multiplets which are chiral under the remnant U($1$) and furthermore are charged under an additional gauge interaction which confines at the scale $\widetilde{\Lambda}_N$. The chiral anomaly of this additional gauge symmetry induces an axion-like potential for the field $\phi_N$. At low scales only the $S$ part of $\phi_N$ survives. Below the confinement scale $\widetilde{\Lambda}_N$, this coupling generates a periodic potential for the relaxion field $\phi$ in $S$ of the form
\begin{equation}
V_N\sim \widetilde{\Lambda}_N^4 \cos \left(\frac{\phi}{2^N f_0}\right) ~.
\end{equation}
where the $2^N$ factor arises because $\phi_N$ is a very small component of $S$. When $N$ is very large, this potential mimics the soft symmetry breaking
term in Eq.~\eqref{eq:relaxionpotential} with
\begin{equation}
 |m_S|^2 \simeq \frac{\widetilde{\Lambda}^4_{N}}{4^N f_\phi^2} ~,
\end{equation}
where we have used $f_\phi \sim f_0$.

The coupling of $S$ with $N$ and $\bar{N}$ can be generated through
the following K\"{a}hler potential terms:
\begin{equation}
 i\frac{\kappa}{{\widetilde M}_N^2} \int d^4\theta N\bar{N} \Xi^*
  \bar{\Xi}^* +{\rm h.c.} ~,
\end{equation}
where $\kappa$ is a constant and ${\widetilde M}_N$ is a UV scale. Since
this operator breaks the chiral symmetry of $N$ and $\bar{N}$, it might
be additionally suppressed by a small factor such as $m_N$. This effect
can be included in the constant $\kappa$. The chiral multiplets $\Xi$ and $\bar{\Xi}$ which are charged under the new strong gauge interaction will confine at $\widetilde{\Lambda}_N$ and the
fermionic components, $\xi$ and $\bar{\xi}$, respectively, condense with $\langle \bar{\xi} \xi \rangle
\simeq \widetilde{\Lambda}_N^3$. Below the confinement scale, these
terms induce the $\Pi_N$--$N$--$\bar{N}$ coupling, which then gives
the $S$--$N$--$\bar{N}$ coupling:\footnote{This also contributes to the
mass term of $N\bar{N}$. }
\begin{equation}
 i\frac{\kappa}{{\widetilde M}_N^2} \int d^2\theta ~
  \widetilde{\Lambda}_N^3e^{\frac{\Pi_N} {f_N}}
N\bar{N}  +{\rm h.c.}
\simeq
\int d^2 \theta \frac{i\kappa  \widetilde{\Lambda}_N^3 }{f_\phi 2^N
{\widetilde M}_N^2} SN\bar{N}
+\text{h.c.}~,
\end{equation}
and thus we obtain
\begin{equation}
 g_S \simeq  \frac{\kappa\widetilde{\Lambda}_N^3 }{2^N f_\phi {\widetilde M}_N^2} \simeq
 \frac{\kappa |m_S| \widetilde{\Lambda}_N }{{\widetilde M}_N^2}~.
\end{equation}
Note in particular that $g_S$ can be substantially smaller than the naive estimate $\sim m_S/f_\phi$.

It is not necessary for the amplitudon, $\sigma$ to be a Nambu-Goldstone field since it does not have a periodic potential. For this reason we keep the same terms as those given in Sec.~\ref{sec:model} for this field, although a similar set up as that considered for ${\cal S}_S$ may be invoked to explain the small shift-symmetry breaking of ${\cal S}_T$.

In summary, by using multi-axion-like fields we have shown that the relaxion potential given in Eqs.~\eqref{eq:relaxionpotential} and \eqref{eq:simpamp} can be derived from a UV model.

\newpage
\bibliographystyle{JHEP}
\bibliography{ref}


\end{document}